\PassOptionsToPackage{hyphens}{url}
\PassOptionsToPackage{obeyspaces}{url}
\PassOptionsToPackage{spaces}{url}
\PassOptionsToPackage{dvipsnames}{xcolor}

\PassOptionsToPackage{hyphens}{url}
\documentclass[letterpaper,twocolumn,10pt]{article}

\usepackage[normalem]{ulem}
\usepackage{soul}

\usepackage{flushend}
\usepackage{balance}
\usepackage{multirow}
\usepackage{graphicx}
\usepackage{algorithm}
\usepackage{algpseudocode}
\usepackage{makecell}
\usepackage{tikz}
\usepackage{environ}
\usepackage{authblk}
\usepackage{ulem}
\usepackage{pdfpages}
\usepackage{amsmath}

\usepackage{caption}

\usepackage{usenix2019_v3}

\usepackage{titlesec}
\titlespacing{\section}{0pt}{6pt plus 2pt minus 2pt}{4pt plus 2pt minus 2pt}
\titlespacing{\subsection}{0pt}{6pt plus 2pt minus 2pt}{4pt plus 2pt minus 2pt}
\titlespacing{\subsubsection}{0pt}{3pt plus 2pt minus 2pt}{2pt plus 1pt minus 1pt}
\titlespacing{\paragraph}{0pt}{\parskip}{-\parskip}

\newcommand{\del}[1]{}

\NewEnviron{delenv}{}

\newcommand{\ignore}[1]{}

\newcommand{\edit}[1]{\textcolor{orange}{#1}}
\newcommand{\editB}[1]{\textcolor{blue}{#1}}

\renewcommand{\edit}[1]{\textcolor{black}{#1}}
\renewcommand{\editB}[1]{\textcolor{black}{#1}}

\begin{document}

\pagenumbering{gobble}

\date{}
\title{Cloudy with a Chance of Cyberattacks:\\Dangling Resources Abuse on Cloud Platforms}

\author[$\S\ddag$]{Jens Frieß}
\author[$\S$*]{Tobias Gattermayer}
\author[$\star$]{Nethanel Gelernter}
\author[$\S\dag$]{Haya Schulmann}
\author[$\S\ddag$*]{Michael Waidner}
\affil[$\S$]{National Research Center for Applied Cybersecurity ATHENE}
\affil[*]{Fraunhofer Institute for Secure Information Technology SIT}
\affil[$\ddag$]{Technische Universität Darmstadt \hspace{7pt} {}$^{\star}$IONIX \hspace{7pt} {}$^{\dag}$Goethe-Universität Frankfurt}

\maketitle

\renewcommand{\figurename}{\footnotesize Figure}
\renewcommand{\tablename}{\footnotesize Table}
\captionsetup[figure]{margin=-2pt,font=footnotesize,labelfont=bf}
\captionsetup[table]{margin=-2pt,font=footnotesize,labelfont=bf}

\thispagestyle{plain} \pagestyle{plain}

\begin{abstract}
Recent works showed that it is feasible to hijack resources on cloud platforms. In such hijacks, attackers can take over released resources that belong to legitimate organizations. It was proposed that adversaries could abuse these resources to carry out attacks against customers of the hijacked services, e.g., through malware distribution. However, to date, no research has confirmed the existence of these attacks. 

We identify, for the first time, real-life hijacks of cloud resources. This yields a number of surprising and important insights. First, contrary to previous assumption that attackers primarily target IP addresses, our findings reveal that the type of resource is not the main consideration in a hijack. Attackers focus on hijacking records that allow them to determine the resource by entering freetext. The costs and overhead of hijacking such records are much lower than those of hijacking IP addresses, which are randomly selected from a large pool. 

Second, identifying hijacks poses a substantial challenge. Monitoring resource changes, e.g., changes in content, is insufficient, since such changes could also be legitimate. Retrospective analysis of digital assets to identify hijacks is also arduous due to the immense volume of data involved and the absence of indicators to search for. To address this challenge, we develop a novel approach that involves analyzing data from diverse sources to effectively differentiate between malicious and legitimate modifications. Our analysis has revealed 20,904 instances of hijacked resources on popular cloud platforms. While some hijacks are short-lived (up to 15 days), $\frac{1}{3}$ persist for more than 65 days. 

We study how attackers abuse the hijacked resources and find that, in contrast to the threats considered in previous work, the majority of the abuse (75\%) is blackhat search engine optimization. We also find fraudulent certificates and stolen cookies. We cluster the abuse resources and abuse content to identify about 1,800 individual attacking infrastructures.

\end{abstract}

\section{Introduction}

Digital resources form the fabric of modern societies. They provide the fundamental platform for digital services and assets, e.g., for financial services, critical infrastructure, government services.
Due to their importance, digital resources pose a lucrative target for attackers. Therefore securing these resources and correctly managing them is crucial for the security of the Internet. Managing resources requires not only creating and configuring them, but also releasing them correctly after they are no longer required. Previous work \cite{liu2016all} showed that when organizations release resources of services that are no longer needed, they often do not purge the infrastructure that was set up for them, creating {\em dangling resources}.

{\bf Dangling records.} Previous work \cite{liu2016all,borgolte2018cloud,alowaisheq2020zombie,squarcina2021can,zhang2023detecting} studied the threat introduced by {\em dangling records}, i.e., Domain Name System (DNS) records that point to resources that were released. The concept of dangling records is related to dangling pointers in programming, which occur when a variable's memory is deallocated. Similarly, DNS records become dangling when domain owners forget to purge the records. For example: a domain owner does not remove a mapping {\small{{\tt foo.com A 1.2.3.4}}} of service {\small{{\tt foo.com}}} to a cloud IP address {\small{{\tt 1.2.3.4}}} from the authoritative DNS server after the resource at {\small{{\tt 1.2.3.4}}} is discontinued and released. Adversaries, which succeed in taking over the released resources that are pointed to by the existing DNS record, can launch attacks against clients that attempt to access the domain. In our example, if an adversary can take over {\small{{\tt 1.2.3.4}}} it can obtain control over all the records that point to that IP address, since all requests to {\small{{\tt foo.com}}} are sent to the adversary. \edit{The attack does not require any extensive capabilities. All that it requires is some way of collecting domain names (e.g., via passiveDNS or Certificate Transparency), checking if the resource is hosted in the cloud and is reachable, and if not, registering the resource through an account with the cloud provider.}

{\bf Research finds multiple dangling records.} In 2016 \cite{liu2016all} analyzed dangling records on cloud and other platforms and the threat that they create for hijacking domains. \cite{liu2016all} found 467 dangling records in top 10K Alexa domains and 52 {\small{\tt .edu}} domains. A follow-up study \cite{borgolte2018cloud} extended the methodology of \cite{liu2016all} for identifying dangling records on cloud platforms and identified over 700,000 dangling DNS records. \cite{squarcina2021can} improved the subdomain enumeration of \cite{liu2016all} and discovered exploitable vulnerabilities in 887 domains. \cite{alowaisheq2020zombie} studied the risk of stale NS records finding 628 hijackable domains.
Recently \cite{zhang2023detecting} developed a hostingChecker, eventually finding 10K vulnerable subdomains. Although dangling records and their threat have been extensively studied in previous work, no research has provided evidence that dangling records are abused for attacks and demonstrated real-life abuses.

{\bf We study real-life abuse of dangling records.} The most remarkable result of our work is the first evidence and analysis of actual, real-life attacks that abuse dangling DNS records. Detecting real-life abuses is hard. The fundamental challenge is detecting malicious vs. legitimate changes in resources. We find that the hijacked resources often do not stand out and even have valid (yet fraudulent) certificates. Approaches that look for changes in the infrastructure or in the content do not work, since changes are often legitimate and happen not only in abused resources. In addition, the huge data volumes involved and lack of known indicators make finding abuses equivalent to looking for a needle in a haystack. We show that the key to finding real-life abuses is a combination of longitudinal data analysis from multiple sources with clustering of changes according to similarities and manual keyword derivation. Applying this approach we derive indicators which enable detection of real-life hijacks. Our longitudinal study of abuses in 12 cloud platforms identified 20904 hijacks that hosted \edit{malicious content}. We detect hijacked domains in 219 Top Level Domains (TLDs) and abuses on popular clouds.

{\bf Selection of hijacked resources is financially motivated.} Previous work measuring dangling records on cloud platforms looked for released IP addresses that were still pointed to by DNS records. In our study we surprisingly find no IP takeovers among real-life abuses of dangling records. The analysis of the abuse cases in our longitudinal dataset shows that the selection of resources by attackers is financially motivated: attackers target dangling resources which can be easily and cost-effectively taken over. These requirements do not apply to IP addresses on popular cloud platforms. Therefore, although the threat of IP address take-over considered in previous work is real, we find that attackers target different resources than previously assumed. We characterize the resources abused by real-life adversaries and explain which factors make them lucrative targets.

We also show that, surprisingly, the most popular abuse of the hijacked resources (75\%) is blackhat Search Engine Optimization (SEO), rather than, e.g., malware distribution, as suggested previously \cite{squarcina2021can}.

{\bf Definition: hijacks \& abuse.} We use the term "hijack" to refer to the appropriation of a (sub)domain name through the re-registration of a released cloud resource pointed to by a dangling DNS record. We use the term ``abuse" to refer to the subsequent use of such a resource for malicious purposes, such as blackhat SEO, clickjacking, phishing, etc. Our definition therefore falls into the type 2 and 3 categories of DNS abuse, defined by \cite{dns-abuse-def-eu}, and subsumes their definitions of ``malicious conduct", ``abusive activity", ``DNS abuse" and ``DNS misuse". We refer to ``blackhat SEO'' as ``SEO", unless being specifically discussed in the context of regular SEO.

{\bf Ethics and notifications.} We initiated a notification campaign and already notified more than 300 organizations of the abuse we found in their resources, which already confirmed the hijacks. Large-scale vulnerability studies pose risks and therefore such research, e.g., \cite{durumeric2013zmap,durumeric2014internet,costin2014large,liu2015cloudy,vandersloot2016towards,kumar2019all,izhikevich2021lzr,alrawi2021circle}, explicitly takes ethics of scans and collected data into account. Due to the sensitive nature of our findings we also take extensive measures to ensure the security of the organizations studied in this work. In our study and data collection we follow the ethical guidelines for network measurements \cite{durumeric2013zmap,partridge2016ethical}, which were also approved by the ethics committee (IRB) in our organization. By following these guidelines we ensure that the equipment of target organizations and cloud platforms is not affected or overloaded, and that the organizations' private data is not compromised. In addition, we conducted a privacy impact assessment with our legal department, which allowed us to conduct the study.
For each organization in our dataset we send at most two HTTP requests per Fully Qualified Domain Name (FQDN) to check an abuse: the first request is for the page itself, and if we cannot establish an abuse with confidence, we send another request for the sitemap. We repeat this data collection on a weekly basis.

{\bf Contributions.} We develop a methodology and use it to find and analyze real-life abuse of dangling records. 

$\triangleright$ {\em Longitudinal comprehensive dataset.} We collect a longitudinal dataset of (sub)domains pointing to deallocated cloud assets, which started with 1,508,273 records and after three years grew to 3,101,992 records.

\indent $\triangleright$ {\em Methodology to detect abuses.} We develop the first methodology that identifies abuse of dangling records. Key to finding abuses are longitudinal data collection from multiple sources and a novel analysis methodology. 
In our dataset we find 20,904 cases of abused dangling records that belong to organizations in multiple sectors.

\indent $\triangleright$ {\em Attackers prefer cheap and easy hijacks.} \cite{borgolte2018cloud} showed it was possible to take over cloud IP addresses pointed to by dangling DNS records. We find that adversaries avoid IP addresses, which are typically randomly allocated from a large pool, and instead target cheap and easy-to-take-over resources.

\indent $\triangleright$ {\em Attacker capabilities.} We show that the abuse that the attacker can launch against the clients of the victim domains is a function of the dangling resource that the attacker takes over.
We develop a model of the attacker capabilities as a function of the dangling resource, extending \cite{squarcina2021can} which only focused on the configuration in the legitimate service.

\indent $\triangleright$ {\em Characterization of abuses.} We find that the main abuse (75\%) is {SEO}. The attackers mostly target domains with established reputation to increase the ranking of their malicious content by search engines. We also identify cookie stealing attacks, fraudulent certificates and malware distribution - and we analyze these attacks. Overall, we find that the hacking groups successfully attacked 31\% of the Fortune 500 companies and 25.4\% of the Global 500 companies, some over long periods of time. Many of the victim organizations were abused more than once, with one even suffering abuse across more than 100 different subdomains.

\indent $\triangleright$ {\em Characterization of attackers.} We develop a methodology for clustering attackers into groups, using content and meta-data on the abuse sites and the infrastructure of the attackers. We identify 1,800 individual attackers infrastructures.

\indent {\bf Organization.} We review related work in Section \ref{sc:works}. In Section \ref{sc:dataset} we develop a methodology for collecting hijacks. In Section \ref{sc:analysis:abuses} we characterize hijacked resources and in Section \ref{sc:abuse} analyze the abuse deployed on hijacked resources. In Section \ref{sc:attribution} we develop methods to cluster the abuse by attacker infrastructure and conclude in Section \ref{sc:conc}.

\section{Related Work}\label{sc:works}

The threat of hijacking domains by taking over released resources, pointed to by dangling (stale) DNS records, was considered in previous work  \cite{liu2016all,borgolte2018cloud,squarcina2021can,zhang2023detecting}. The idea is that the attacker attempts to get assigned a recently released resource, therefore taking over the domain which points to that resource. \cite{liu2016all} showed that it was practical for an attacker to obtain the desired IP address from the cloud pool by repeatedly allocating and releasing IP addresses. The authors scanned cloud IP addresses to find dangling DNS records from their dataset of domains using Zmap \cite{durumeric2013zmap} and found hundreds of dangling records on cloud platforms and on top 10K Alexa domains. \cite{borgolte2018cloud} extended the dataset of \cite{liu2016all} and collected 130M domains that point to IP addresses in cloud platforms. \cite{borgolte2018cloud} found that over 700,000 domains point to cloud IP addresses that were free and hence vulnerable to domain takeover attacks. \cite{borgolte2018cloud} also estimated that it would be economically practical for attackers to obtain a target IP address from the cloud pool. 
\cite{squarcina2021can} further improved the subdomain enumeration and their analysis of deprovisioned cloud instances yielded 13,532 potentially vulnerable domains with dangling records. \cite{alowaisheq2020zombie} analyzed dangling NS records and found 628 hijackable domains.
Recently, \cite{zhang2023detecting} developed an automated framework for detecting dangling records by reconstructing DNS resolution chains and found 10K subdomains among the top 1M Tranco domains with dangling records.
Our research augments the previous work in the following aspects:

{\em Targeted dangling records.} While \cite{squarcina2021can} suggested that the likelihood of a domain being vulnerable is directly related to the number of subdomains it has, we show that the ease of taking over the dangling DNS record and the reputation of the target domain define the likelihood of an attack.

{\em Analysis of the attack surface.} To measure the prevalence of the dangling records previous work sent liveness probes to IP addresses in cloud IP ranges to determine if the IP addresses were allocated and in use. \cite{liu2016all} checked TCP ports 80 and 443 and TCP/UDP port 53, \cite{borgolte2018cloud} sent ICMP pings and TCP/UDP requests to 36 common TCP/UDP ports, and \cite{squarcina2021can} scanned 148 TCP/UDP ports. Records that pointed to IP addresses that did not respond on any ports were classified as dangling.

Due to virtual hosting, TCP/UDP and ICMP pings do not accurately reflect the availability of a (web)service. Reaching a virtually hosted service requires connecting on the application layer, rather than the transport layer, in order to traverse the forwarding logic in the webserver. To illustrate this we compare ICMP, TCP and HTTP requests using our dataset of cloud-hosted, hijacked domains. Using ICMP pings we receive responses from 72\% of the domains in the cloud. On TCP ports 80/443 we receive responses from 93\% of the domains. An HTTP request to the respective FQDN results in 89\% responsive domains.
The results indicate that ICMP pings tend to overestimate unresponsiveness, and therefore vulnerability of services by around 20\%, whereas TCP pings tend to underestimate by around 4\%, compared to HTTP requests to the actual FQDNs.
Therefore, to accurately check domain liveness we download HTML files via HTTP/S from the \textit{services} rather than simply probing the ports on the target \textit{servers}, therefore capturing the availability of each individual FQDN, regardless of virtual hosting.

{\em Abuse of dangling records.} In contrast to previous work that measured the prevalence of dangling records, in our work we identify and analyze real-life abuse of the dangling records. We analyze the resources that the attackers take over, characterize the target domains and the abuses of the dangling resources for attacks. Previous work proposed that dangling records could be exploited for stealing cookies, issuing fraudulent certificates, loading malware or authentication bypass. In our work we show that the most common abuse of dangling records is for SEO, which was not considered in prior work.

\section{Collection of Abused Resources}\label{sc:dataset}
In January 2020 we started monitoring 1.5 million cloud assets in use by large organizations. After a couple of months we noticed that, after a time period of being inactive, domains of large organizations became active again. However, all those pages that became active after a period of inactivity, shared something in common: they had similar error pages, in different languages, reporting that the website was under maintenance; an interested reader is referred to Figure \ref{abuse-2020-05-censured} in the Appendix for one such example in English on a Fortune 500 company domain. Further analysis of these domains revealed that behind the web pages with the error messages were thousands of other pages.
In all those cases the hackers gained control over the domains by taking over abandoned resources on cloud platforms.
We identified such abuse patterns in hundreds of domains, belonging to governments, universities and enterprises worldwide. Periodically adding more organizations to our list, we were identifying more hijacked domains. In June 2020 we were already tracking more than two million domains, over multiple cloud assets, most of them in Azure and AWS (see Table \ref{tab:monitored} for full list of asset types). In this section we describe our three-year data collection methodology and the resulting dataset of assets. An overview of this collection process is shown in the top left (blue) of Figure \ref{fig:data-flow}. This data forms the basis for the analyses, shown in the top right (green), which are covered in Sections \ref{sc:analysis:abuses}, \ref{sc:abuse} and \ref{sc:attribution}.

\subsection{Dataset}
Our \editB{initial search space} contains domains across a number of sectors: a list of 2M government domains\footnote{\texttt{.gov} filtered from \path{https://domainsproject.org/}}, Fortune 1000\footnote{\path{https://fortune.com/analytics/fortune-1000}} and Global 500 enterprise domains\footnote{\path{https://fortune.com/ranking/global500/}}, and 1M-top Alexa domains\footnote{\path{https://www.kaggle.com/datasets/cheedcheed/top1m}}. We also use a list of 9,933 university domains\footnote{\path{https://github.com/Hipo/university-domains-list}}. \editB{These domains serve as candidates for finding potential hijacks. We do not sanitize these further, as inaccessible domains are automatically removed as part of our search methodology.} Using \edit{the FarSight} passive DNS service, with global sources across all continents, we also \edit{discover} all subdomains observed for these domains. From this initial list of known high-profile (sub)domains we determine the subset that points to cloud assets, resulting in a list of $1,508,273$ (sub)domains, which constitutes the dataset we study in this work.

\begin{algorithm}[t!]
\scriptsize
\caption{{\footnotesize{Collection of Cloud-pointing FQDNs.}}}\label{alg:forward-collection}
\begin{algorithmic}[1]
\Function{collect\_fqdns}{$fqdns$, $cloud\_suffixes$, $cloud\_IPs$}
    \State $fqdns\_to\_analyze \gets \{~\}$
    \For{$fqdn$ \textbf{in} $fqdns$}
        \State $A\_results,~CNAME\_results \gets$ DNS\_A\_query($fqdn$)
        \For{$CNAME$ \textbf{in} $CNAME\_results$}
            \If{$CNAME$.ends\_with\_any($cloud\_suffixes$)}
                \State $fqdns\_to\_analyze$.add($fqdn$)
            \EndIf
        \EndFor
        \For{$IP$ \textbf{in} $A\_results$}
            \If{$IP$ \textbf{in} $cloud\_IPs$}
                \State $fqdns\_to\_analyze$.add($fqdn$)
            \EndIf
        \EndFor
    \EndFor
    \State \textbf{return} $fqdns\_to\_analyze$
\EndFunction
\end{algorithmic}
\end{algorithm}

The pseudocode of our methodology is described in Algorithm \ref{alg:forward-collection}. We select domains and subdomains that have a CNAME DNS record referencing an FQDN with a known cloud suffix (e.g., \texttt{*.azurewebsites.net}, \texttt{*.amazonaws.com}); we provide a list of the known cloud suffixes relevant for our research in Appendix \ref{sc:cloud:suffixes}. For domains without CNAMEs we check if one of its IPs falls within a subnet used for cloud hosting. This is often the case with domains hosted on AWS S3 buckets or dedicated VM servers, which refer to these servers using A records rather than CNAME. \edit{Subnet information is published by the cloud providers (see Appendix \ref{sc:cloud:suffixes}). By downloading these regulary we ensure that we are using up-to-date information.}

\begin{figure}[t!]
    \centering
    \vspace{-10pt}
    \includegraphics[width=0.8\columnwidth]{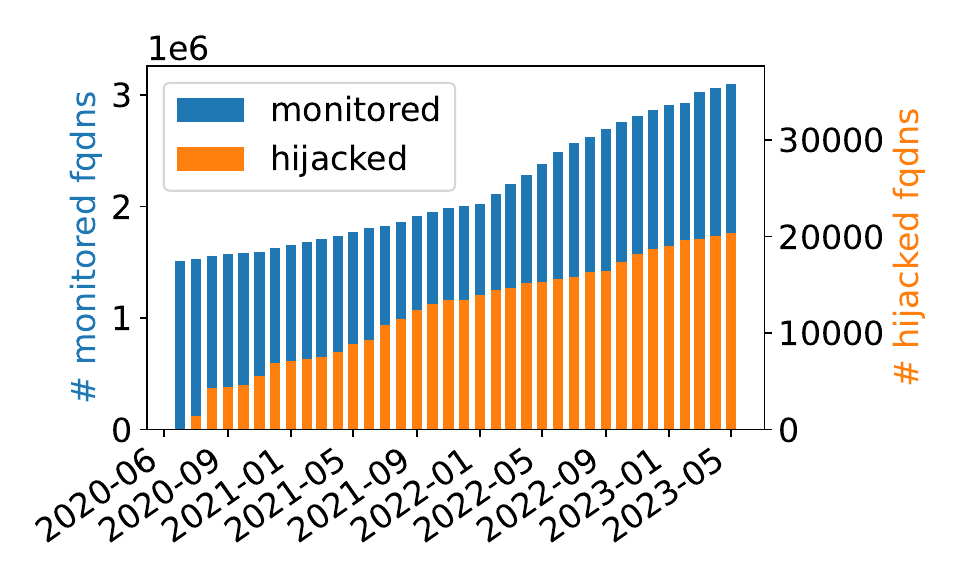}
    \vspace{-15pt}
    \caption{Monitored vs. hijacked cloud-hosted domains over time.}
    \vspace{-10pt}
    \label{fig:domains-monitored}
\end{figure}

\indent Over the 3 year period, we kept updating the list by consuming a commercial feed of FQDNs for the enterprises in our dataset (in Section \ref{sc:dataset}). Within that time, we filtered through more than 87,000,000 non-NXDOMAIN (i.e., with at least one DNS record) domains and subdomains, resulting in a doubling of our initial list to 3,101,992 monitored FQDNs.
Figure \ref{fig:domains-monitored} shows this monthly increase of monitored cloud-hosted FQDNs, overlayed with the cumulative number of abuses seen until that time. We collected the data and monitored the changes in DNS records and in HTML files for over 3 years.

\subsection{Detection of Abused Domains}

Finding abuses is hard, hence it is not surprising that so far no analysis of dangling records has found actual abuses. The main problem is that changes in DNS records and websites are often legitimate, and without knowing what malicious content to look for, finding abuse is virtually impossible.
To determine whether assets were abused, we track changes of site content. To accomplish this we take regular samples of each site, downloading the index HTML as well as the sitemap stored in a database. \edit{By comparing these snapshots, including changes to DNS, HTTP response, sitemap (e.g., size changes of 100KB), language changes, and keywords, differences can be detected. The components in our analysis are illustrated in Appendix, Figure \ref{fig:data-flow}).}

{\bf Signatures.} Having identified content changes in groups of assets within a short time frame, we manually inspect the new content \edit{to ensure no false positives when notifying affected organizations. We then create signatures by automatically extracting keywords from the index page and other pages, as well as sitemap features, JavaScript and other loaded objects. We validate that these are shared across multiple abuse pages and finally test these against a large dataset of benign assets to ensure they do not yield false positives; those that do are discarded. The benign assets are also assembled from Alexa websites, Fortune 500 sites and university sites, ensuring cross-sector representation, and verified to not contain malicious content.} Once created, these signatures can be used to detect similar malicious changes on other domains we monitor. Examples of the signatures we create include:

\vspace{5pt}
{\scriptsize{
\begin{verbatim}
(1) index page includes "Comming soon ..." 
//"Comming" is written with a typo instead of "Coming".
(2) index page refers to 3 other pages in a specific 
structure; the referred pages are written to create a 
new window on each click.
(3) loading a particular "popunder.js" script
(4) sitemap with several thousand pages (> 5 MB); 
each page is structured similarly; consistent random name 
generation
(5) New sitemap or 100KB increase in sitemap size.
(6) Language change. 
(7) Keywords related to content uploaded by the attackers
\end{verbatim}
}}
\vspace{5pt}

\begin{figure}[h!]
\vspace{-5pt}
    \centering
    \begin{minipage}[t]{0.46\columnwidth}
        \centering
        \includegraphics[width=\textwidth,trim=0cm 1cm 0cm 1.5cm, clip]{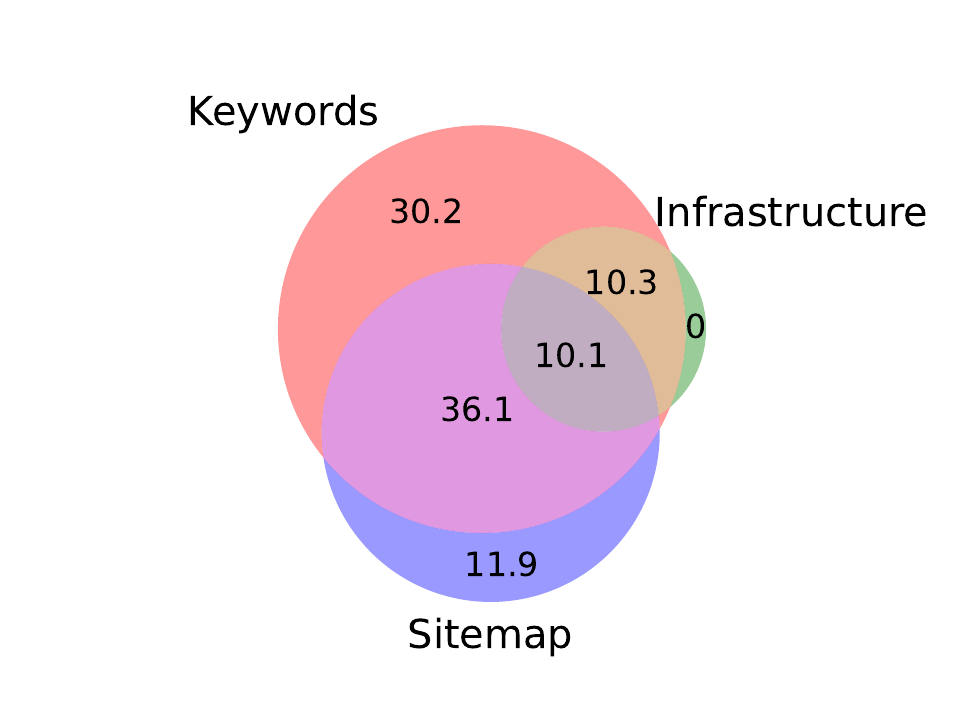}
        \vspace{-10pt}
        \caption{\% of detected hijacks with extracted signatures by type.}
        \label{fig:indicator-venn}
    \end{minipage}
    \hfill
    \begin{minipage}[t]{0.46\columnwidth}
        \includegraphics[width=\textwidth,trim=0cm 1cm 0cm 1.5cm, clip]{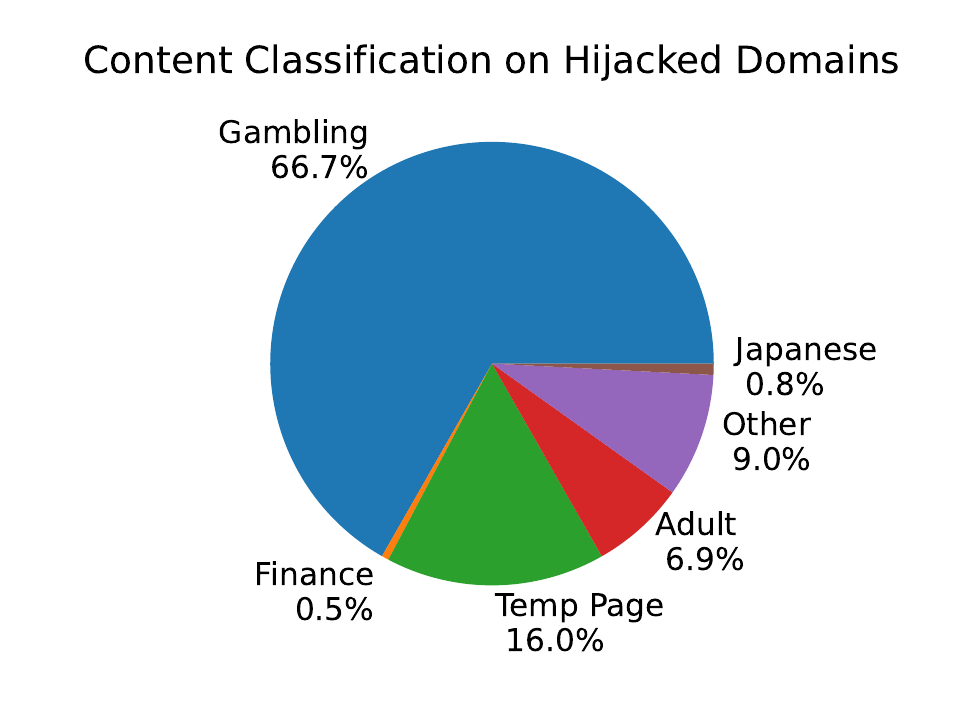}
        \vspace{-10pt}
        \caption{Content classification on hijacked domains.}
        \label{fig:content-classification}
    \end{minipage}
    \vspace{-5pt}
\end{figure}

We validated the changes manually on the abused assets. Figure \ref{fig:indicator-venn} shows what percentage of domains in our dataset match different types of indicators or combinations thereof. For example, some domains (30.2\%) can be identified with just keywords (such as those shown in Table \ref{tab:keyword-counts}), whereas others (10.1\%) require using keywords as well as attacker infrastructure-related indicators, such as hyperlinks or scripts and images loaded from other domains, as differentiating features. \edit{If the required features are present on the site, the signature matches and the domain is classified as abused.}

In our analysis we find that the page contents, i.e., keywords, are the most telling indicators of abused assets, whereas infrastructure-related indicators are only useful in combination with keywords or sitemap features. Looking at the sitemap and keywords in combination is the most effective, identifying an additional 36.1\% of abused assets in our dataset, compared to just keywords.

\textbf{Keywords.} We extracted 56,946 keywords with an average keyword count of 2.72 to classify index HTML files as abused. Table \ref{tab:classification-keyword-counts} shows the top keywords found, with gambling and adult content as the major sources. Figure \ref{fig:content-classification} shows the full distribution of topics found based on keywords. We list the most popular keywords in the Appendix, Figure \ref{fig:keyword-table}.

\begin{table}[H]
\renewcommand{\arraystretch}{0.7}
\centering
\scriptsize
  \resizebox{\columnwidth}{!}{
        \begin{tabular}{c|l|c||c|l|c}
        \textbf{\#} & \textbf{Keyword} & \textbf{Count} & \textbf{\#} & \textbf{Keyword} & \textbf{Count}  \\
        \hline
        1 & \textit{HTML Snippet} & 4615 & 2 & \textit{HTML Snippet} & 4288 \\ 
        \hline       
        3 & \textit{HTML Snippet} & 4199 & 4 & sex & 3257 \\
        \hline
        5 & daftar (list) & 2930 & 6 & porn & 2786 \\
        \hline        
        7 & situs judi (gambling sites)& 2611 & 8 & \textit{HTML Snippet} & 2193 \\
        \hline
        9 & gacor (hot streak) & 2048 & 10 & [j]udi slot online (gambling online) & 1892 \\
        \hline
        11 & situs slot (slot/gambling sites) & 1880 & 12 & slot gacor (hot slot machine) & 1564 \\
        \end{tabular}
    }
    \vspace{-5pt}
  \caption{Top 12 keywords for index.html classification.}
  \vspace{-5pt}
  \label{tab:classification-keyword-counts}
\end{table}

{\bf Abuse dataset.} \edit{After 3 years of monitoring, we detect 17,698 unique, abused FQDNs (where 1,565 are Second-Level Domains (SLDs), Figure \ref{fig:sldVsSubdomains}) across 11,924 unique SLDs and 218 affected TLDs (see Appendix Table \ref{table:tld_counts} for top 12). These abused domain names point to 15,248 unique CNAMEs (11,654 unique IP addresses). This dataset forms the basis for our subsequent analyses. We notified $>300$ affected organizations, which confirmed the abuse.}

\begin{figure}[H]
    \centering
    \includegraphics[width=1.0\columnwidth]{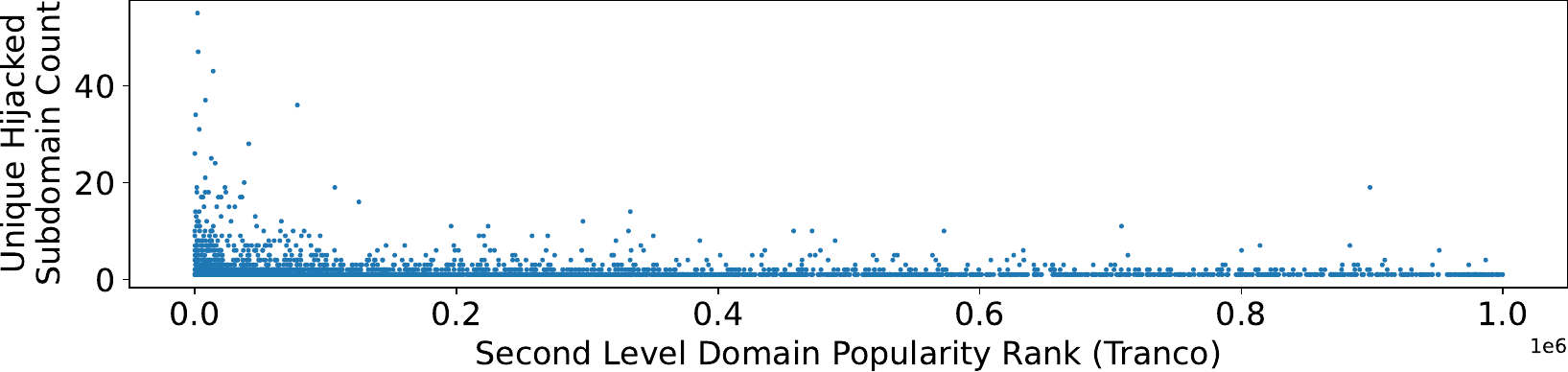}
    \caption{Rank of SLDs and associated hijacked subdomain counts.}
    \label{fig:sld-tranco}
\end{figure}

\edit{The Tranco list is a research-focused ranking of the top websites based on their popularity and stability over time\cite{LePochat2019}. We find 7,049 of 17,698 (39.8\%) unique hijacked FQDNs on SLDs included in the Tranco domain list. On average, every tranco-ranked SLD has 2 (1.89) hijacked subdomains. The rank and corresponding unique hijacked subdomain count are illustrated in Figure \ref{fig:sld-tranco}.}

\begin{figure}[H]
\centering
\vspace{-5pt}
\setlength{\abovecaptionskip}{-1pt}
\includegraphics[width=1.0\linewidth]{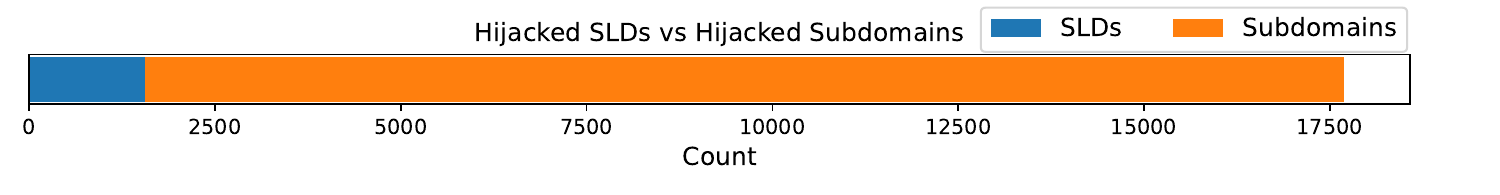}
\vspace{-10pt}
\caption{Abused second level domains and sub-domains.}
\label{fig:sldVsSubdomains}
\vspace{-10pt}
\end{figure}

\begin{figure}[t!]
    \centering
    \includegraphics[width=\columnwidth]{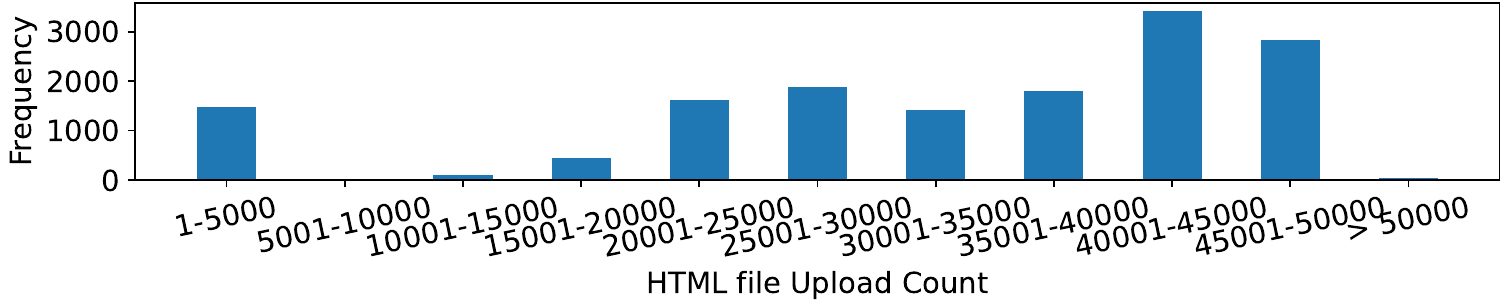}
    \vspace{-15pt}
    \caption{HTML Files uploaded to each abused site}
    \vspace{-10pt}
    \label{fig:sitemap-file-counts}
\end{figure}

\indent {\bf Abuse data volume.} We collected 54,325 index HTML files classified as abused content with an average file size of 52.4kB, some of them for the same assets (with changes made by the hackers).
We also collected 15,482 sitemaps and analyzed them for the quantity of total malicious HTML content uploaded. Figure \ref{fig:sitemap-file-counts} shows a histogram of the number of HTML files uploaded, grouped into bins of 5,000. The number of files ranges from 2 to 144,349 HTML files per site\edit{, with the clear majority of sites containing many thousands of pages}. Abusers uploaded a total of nearly 500M (492,489,492) files with an estimated size of \edit{24TB} (25,806,449,380.8kB) and an average of 31,810 HTML files per site.

\begin{figure*}[t!]
  \centering
  \begin{minipage}[t]{0.3\linewidth}
    \centering
    \includegraphics[width=\linewidth,trim=0cm 0cm 0cm 0.98cm, clip]{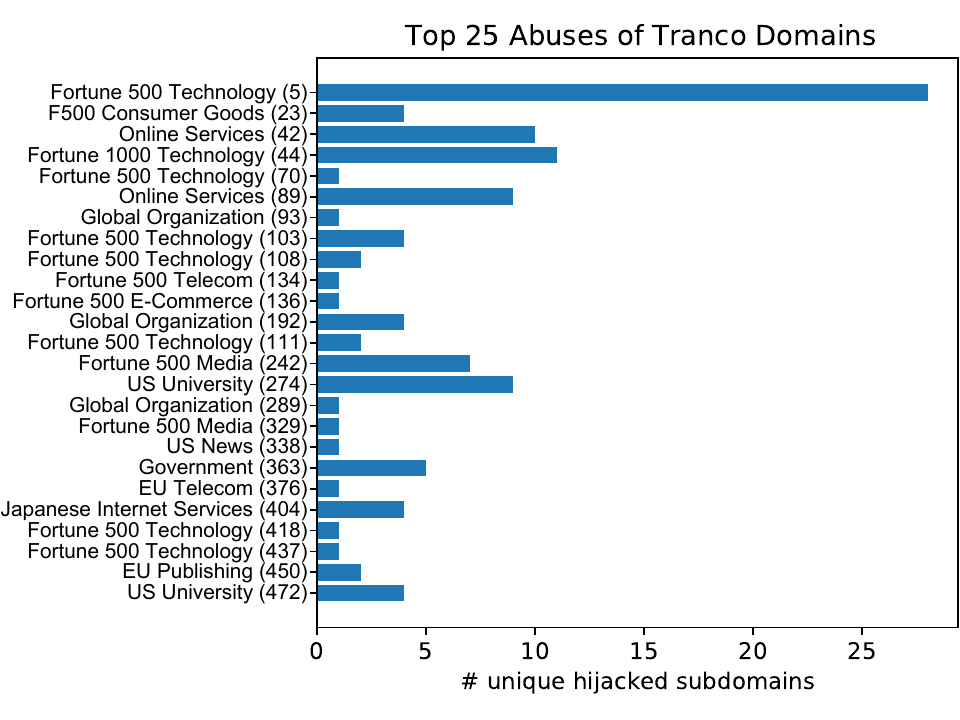}
    \vspace{-20pt}
\caption{Abuse in top 1M Tranco Domains.}
\label{fig:top25-tranco-abuses}
  \end{minipage}
  \hfill
  \begin{minipage}[t]{0.3\textwidth}
    \includegraphics[width=\linewidth,trim=0cm 0cm 0cm 0.98cm, clip]{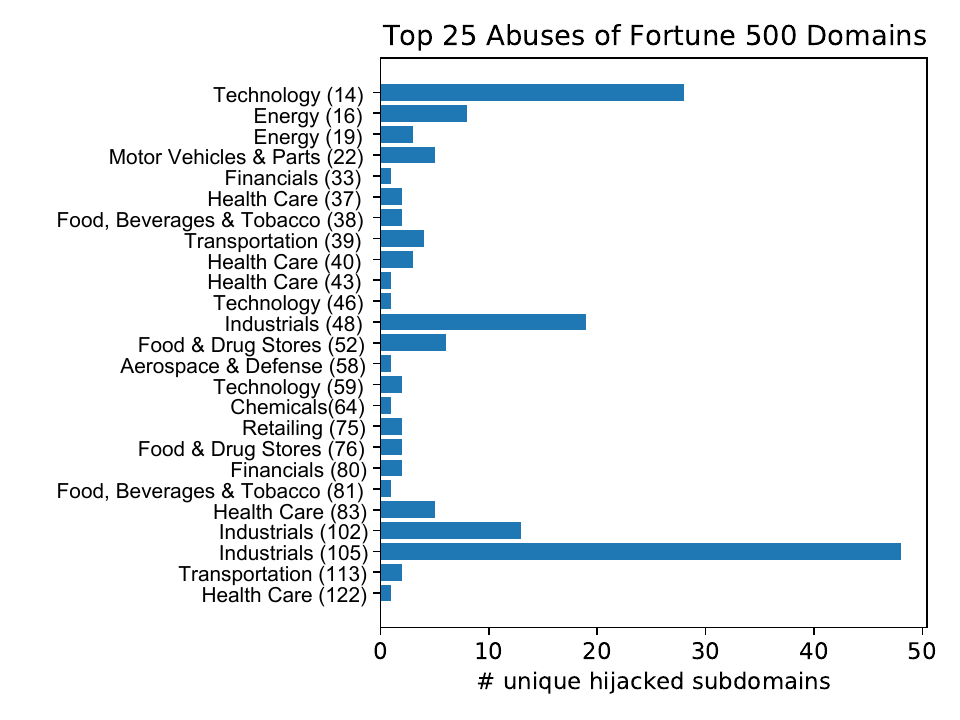}
    \vspace{-20pt}
\caption{Abuse in Fortune 500 companies.}
\label{fig:top25-fortune500-abuses}
  \end{minipage}
  \hfill
  \begin{minipage}[t]{0.3\textwidth}
    \includegraphics[width=\linewidth,trim=0cm 0cm 0cm 0.98cm, clip]{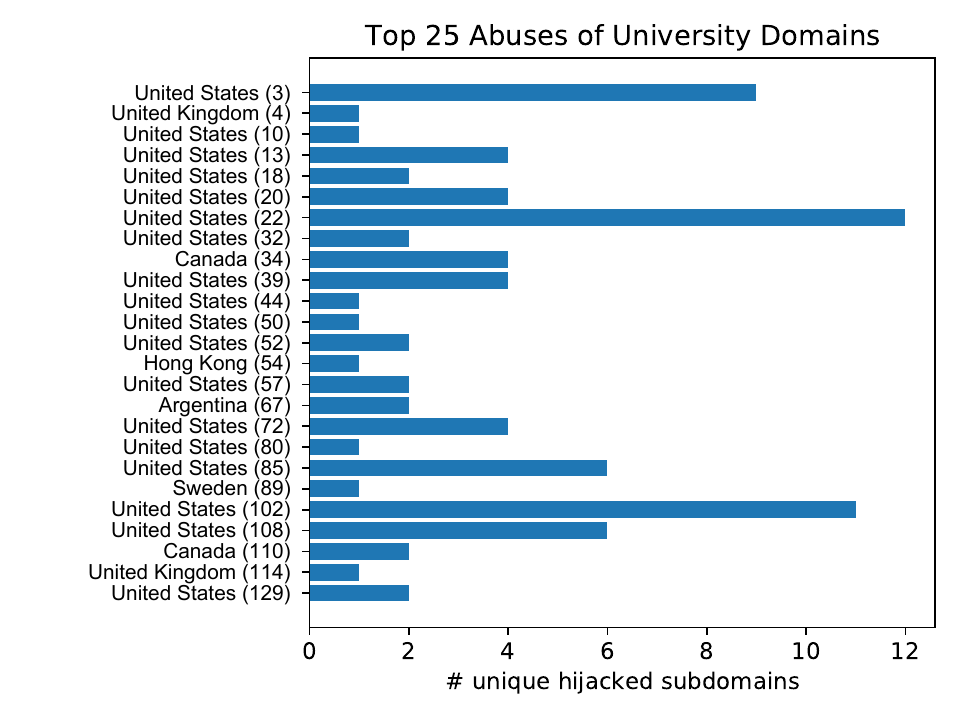}
    \vspace{-20pt}
\caption{Abuse in QS-ranked universities.}
\label{fig:top25-uni-abuses}
  \end{minipage}
  \vspace{-10pt}
\end{figure*}

\indent {\bf Ruling out benign changes.} There are cases in which changes in content can be legitimate, e.g., changes in parked domains, that display commercial HTML content that changes collectively over time. To rule out legitimate changes applied by registrars that manage multiple domains, we analyze information of the registrars and the owners of the domains. Specifically, we check that clusters of domains with identical changes in content, have different owners and registrars.

\begin{figure}[H]
\vspace{-10pt}
    \centering
    \includegraphics[width=0.8\columnwidth]{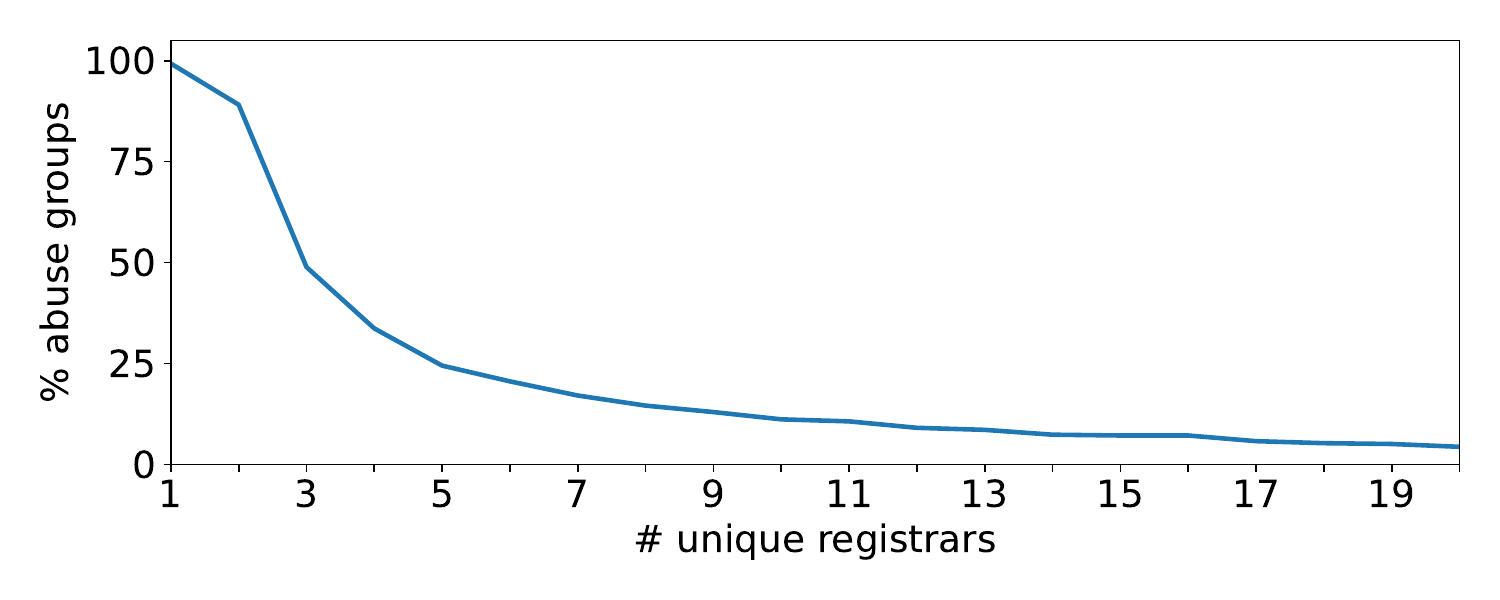}
    \vspace{-7pt}
    \caption{\% abuse clusters (grouped by keywords) spanning $\geq$ X registrars.}
    \label{fig:reg-dist}
\end{figure}

\indent To do this we aggregate domains into clusters based on the keywords extracted from their web pages. Identical keyword lists indicate the same page content. By matching the second-level domains to their respective registrars, we determine the set of unique registrars for each cluster of domains with the same content-changes. We then plot the percentage of these clusters, with at least two domains, by the number of unique registrars observed. This is shown in Figure \ref{fig:reg-dist}.

Our analysis yielded that in 89\% of the cases, where the same change is detected on at least 2 domains, these changes  span 2 or more different registrars and owners. In 33\% of cases, the changes occur across domains owned by 4 or more registrars; see distribution in Figure \ref{fig:reg-dist}. This result demonstrates that identical changes in clusters of domains are not made by the registrars, since the domains typically have different registrars.

\section{Analysis of Abused Resources}\label{sc:analysis:abuses}
We describe resources of organizations that were abused and identify which cloud platforms host abused resources. 
\edit{To confirm a hijack we have downloaded and manually examined malicious content on hijacked domains. The fact that the same content is seen across unrelated, independent domains and that the content is unrelated to the topics of these domains provides an evidence of abuse by a third party. We also analyzed abuse showing that the content hosted on those hijacked resources are malicious in Section \ref{sc:abuse}.}

\subsection{Abused Organisations}

\textbf{Popular domains.} Among the abused domains we find 8,432 popular websites from the Tranco list \cite{pochat2018tranco} (\edit{top 25 shown in} Figure \ref{fig:top25-tranco-abuses}).

\begin{figure}[H]
  \centering
  \vspace{-5pt}
  \begin{minipage}[t]{0.23\textwidth}
    \centering
    \includegraphics[width=\linewidth,trim=0cm 0cm 0cm 1.3cm, clip]{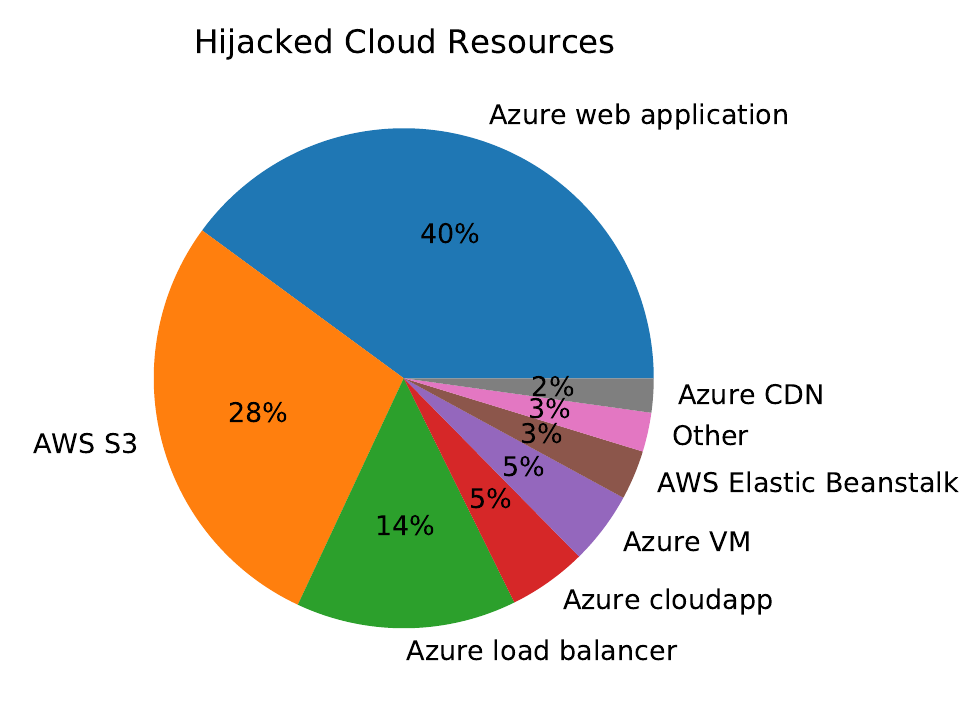}
    \vspace{-25pt}
\caption{{\footnotesize Cloud resources.}}
\label{fig:cloud-res-pie}
  \end{minipage}
  \hfill
  \begin{minipage}[t]{0.23\textwidth}
    \includegraphics[width=\linewidth,trim=0cm 0cm 0cm 1.3cm, clip]{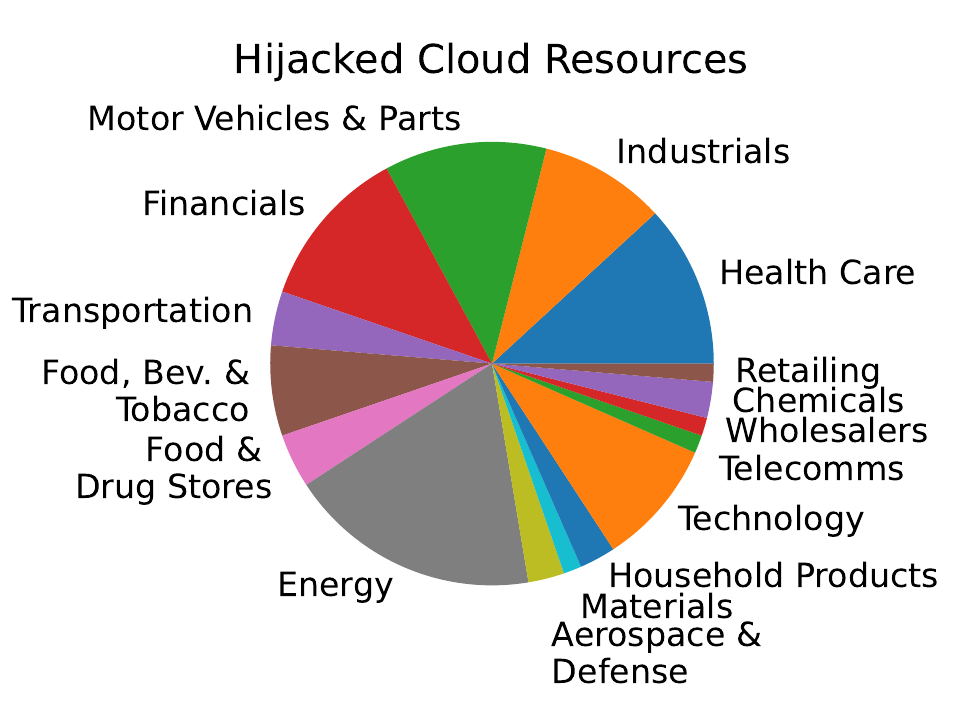}
    \vspace{-25pt}
\caption{{\footnotesize Abused content by sector.}}
\label{fig:abused-by-sector}
  \end{minipage}
\vspace{-10pt}
\end{figure}

\textbf{Enterprises.} We find abuse in 31\% of the Fortune 500 companies\textsuperscript{2} and 25.4\% of the Global 500 companies. Comparing these two lists suggests that the attackers focused on Western countries.
The 25 highest-ranking Fortune 500 enterprises abused are shown in Figure \ref{fig:top25-fortune500-abuses}. Many of the companies were abused more than once, hosting fraudulent content on more than one subdomain at some point in time. Figure \ref{fig:abused-by-sector} shows that the Industrial, Energy and Motor Vehicle sectors have the highest volume of \edit{hijacks, but overall the abuse is widespread rather than localized to any one sector}.

\textbf{Universities.} We find hijacks of university domains worldwide\textsuperscript{5} (\edit{top 25 shown in }Figure \ref{fig:top25-uni-abuses}).
Between May 2020 and 2023 we found 264 abused subdomains in universities globally. University domains have good reputations and are therefore a desirable target for promoting fraudulent content.

\subsection{Abused Providers \& Resource Types}
Hackers exploit a variety of cloud platforms to claim dangling records. \edit{The different services provided by each platform can be determined by the respective cloud domain suffix.} Figure \ref{fig:cloud-res-pie} and Table \ref{tab:monitored} show, that Microsoft Azure Cloud services are hosting more than half of the content, followed by AWS S3 static hosting and AWS Elastic Beanstalk, which together make up $\frac{1}{3}$. All other cloud providers only account for small fractions. A 2016 study showed substantially fewer dangling records pointing to Azure than AWS \cite{liu2016all}.

\begin{table}[ht]
\scriptsize
\renewcommand{\arraystretch}{0.7}
    \centering
        \begin{tabular}{l|r|r|r}
            \textbf{Cloud Resource} & \textbf{\# Monitored} & \textbf{\# Abuses} & \textbf{\% Abuses} \\
            \hline
            Azure Web Application & 690,779 & 8,347 & 1.21 \\
            Azure VM & 565,684 & 983 & 0.17\\
            Azure Blob & 20,389 & - & - \\
            AWS Elasticbeanstalk & 138,523 & 668 & 0.48\\
            Azure Traffic Manager & 140,183 & 2,980 & 0.21\\
            Azure Cloud Service & 299,494 & 1,060 & 0.35\\
            Azure API & 17,100 & - & - \\
            Azure FrontDoor & 14,183 & - & - \\
            Heroku App & 30,532 & 146 & 0.48\\
            Azure CDN & 37,360 & 461 & 1.23\\
            Azure Service Bus & 10,152 & - & - \\
            AWS S3 & 1,137,613 & 5,876 & 0.52
        \end{tabular}
    \vspace{-8pt}
    \caption{Abused cloud services among domains monitored.}
    \vspace{-5pt}
    \label{tab:monitored}
\end{table}

\begin{figure}[t!]
    \centering
    \includegraphics[width=1.0\columnwidth]{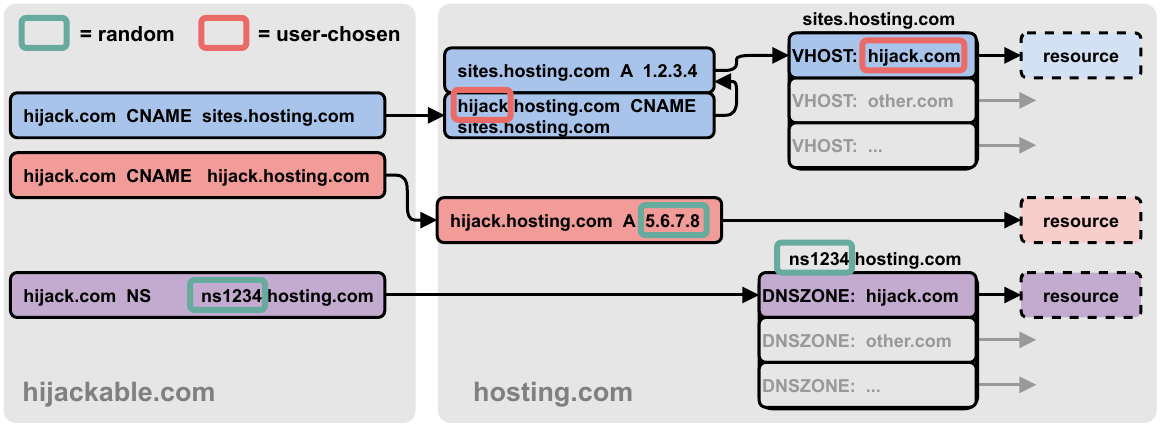}
    \caption{Hijack types: dashed lines are released resources an attacker could re-register to take over the routing and/or content of \texttt{hijackable.com}.}
    \label{fig:hijack-types}
\end{figure}

\subsection{The Problem of User-Nameable Resources}
\begin{figure}[b!]
    \centering
    \vspace{-10pt}
    \includegraphics[width=0.9\columnwidth]{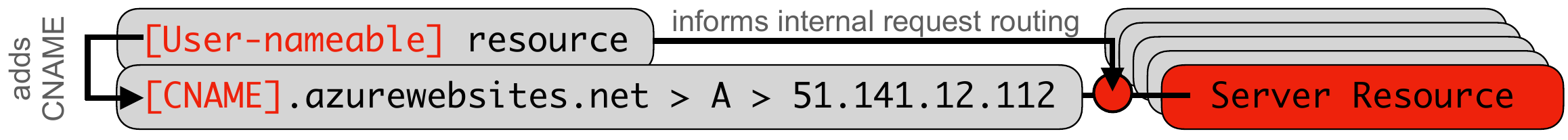}
    \vspace{-5pt}
    \caption{User-nameable resource informs CNAME and internal routing.}
    \label{fig:cloud-routing}
\end{figure}

We find that the common denominator among the hijacks is \textit{user-nameable} resources. As shown in Table \ref{tab:free-text-entry}, all hijacks we discovered exploited cloud resources that allow free choice of a text-based identifier (blue in Figure \ref{fig:hijack-types}), enabling easy re-registration by an attacker.\\
\indent {\bf CNAMEs \& internal routing.} Such user-chosen names are found at the DNS-level and the virtual hosting-level to route requests to the appropriate resource. At the DNS-level, these names are used in resource's domain, auto-generated by the cloud provider to resolve to the resource's IP address. For example, choosing the name \texttt{example} for an Azure website results in the subdomain \texttt{example.azurewebsites.net} being generated, resolving to the IP of one of the Azure servers. The server also configures its virtual hosting layer to route incoming requests for \texttt{example.azurewebsites.net} to the \texttt{example} resource (Figure \ref{fig:cloud-routing}).\\
\indent Cloud providers typically also allow configuration of custom domains as aliases, so that incoming requests for, e.g., \texttt{example.com}, are also directed to the \texttt{example} resource. The customer can then conveniently set up a CNAME record directing \texttt{example.com} to \texttt{example.azurewebsites.net}, allowing the \texttt{example} resource to be resolved through \texttt{example.com} at the DNS level.\\
\indent \textbf{Deterministic re-registration.} However, this system poses two problems. First, the resource name chosen by the legitimate user is publicly visible through the DNS record of the auto-generated CNAME. Second, this resource name can be re-registered by an attacker. An attacker who has found a DNS record pointing at \texttt{example.azurewebsites.net} can check if the \texttt{example} resource still exists and, if not, register this specific resource. As long as the CNAME record linking \texttt{example.com} to \texttt{example.azurewebsites.net} is never purged, requests for \texttt{example.com} (and any other domain in a CNAME chain to \texttt{example.azurewebsites.net}) will be hijacked by the attacker.\\
\indent {\bf Randomized identifiers.} One effective mitigation of these hijacks is to randomly generate the resource names, because then an attacker is not able to deterministically replicate a target resource. This would maintain the convenience of linking custom domains to the resource through a CNAME record, but provide comparable security to the random IP assignment used for cloud servers with dedicated public IPs.\\
\indent Cloud resources with dedicated IP addresses are assigned their IP at random from a pool available to the cloud provider (red in Figure \ref{fig:hijack-types}). Similarly, cloud providers who offer DNS hosting distribute user-created DNS zones randomly across a range of nameservers (purple in Figure \ref{fig:hijack-types}). When a dangling A or NS record, respectively, points at the IP of such a resource, attackers must register a similar resource repeatedly in the hope of being assigned the desired IP.\\
\indent Previous work \cite{liu2016all,borgolte2018cloud} showed strategies to do this effectively, but it is still a probabilistic technique, which, according to our data, attackers do not pursue. Since our collection methodology (see Algorithm \ref{alg:forward-collection}) also takes A records pointing to cloud IPs into account, such takeovers of specific IPs would also be captured in our dataset. However, we find no instances of such takeovers in our dataset, suggesting that it is not worth the effort compared to the deterministic approach possible with user-nameable resources. This is further underlined by the absence of abused Google Cloud-hosted domains, which are assigned a random subdomain, allowing no user input.

\begin{table}[t!]
\scriptsize
\renewcommand{\arraystretch}{0.7}
  \resizebox{\columnwidth}{!}{
        \begin{tabular}{c|l|l|l|r}
          \textbf{Provider} & \textbf{Configurable Sub domain name} & \textbf{Function} & \textbf{Record} & \textbf{Abuses}\hspace{0cm} \\
          \hline
              Azure & [freetext].azurewebsites.net & Web App  & CNAME & 6,288\hspace{0cm}  \\
              Azure & [freetext].trafficmanager.net & Traffic Router & CNAME  & 1,468\hspace{0cm}  \\
              Azure & [freetext].cloudapp.net (legacy naming) & VM & CNAME  & 1,037\hspace{0cm}   \\
              Azure & [freetext].azureedge.net  & CDN & CNAME  & 830\hspace{0cm}  \\ 
              Azure & [freetext].REGION.cloudapp.azure.com  & VM  & CNAME  & 928\hspace{0cm}  \\
              Azure & [freetext].sip.azurewebsites.windows.net & Web App & CNAME  & 223\hspace{0cm}  \\
          \hline
              AWS & [freetext].s3-website.REGION.amazonaws.com & Static Hosting & CNAME  & 2,227\hspace{0cm}  \\     
              AWS & [freetext].REGION.elasticbeanstalk.com  &  Orchestration & CNAME   & 555\hspace{0cm}  \\ 
          \hline
              Heroku & [freetext].herokuapp.com & Web App & CNAME  & 139\hspace{0cm} \\
          \hline 
              Pantheon & [ test- | dev- | live- ][freetext].pantheonsite.io & CMS & CNAME & 50\hspace{0cm} \\
           \hline
                Netlify & [freetext].netlify.app & Web App & CNAME & 14\hspace{0cm} \\

        \end{tabular}
    }
        \vspace{-5pt}
  \caption{Abused resources on cloud platforms with free text entry.}
  \label{tab:free-text-entry}
\end{table}

\subsection{Abuse Duration}
\begin{figure}[t!]
  \centering
  \begin{minipage}[t]{0.23\textwidth}
    \centering
    \includegraphics[width=\linewidth]{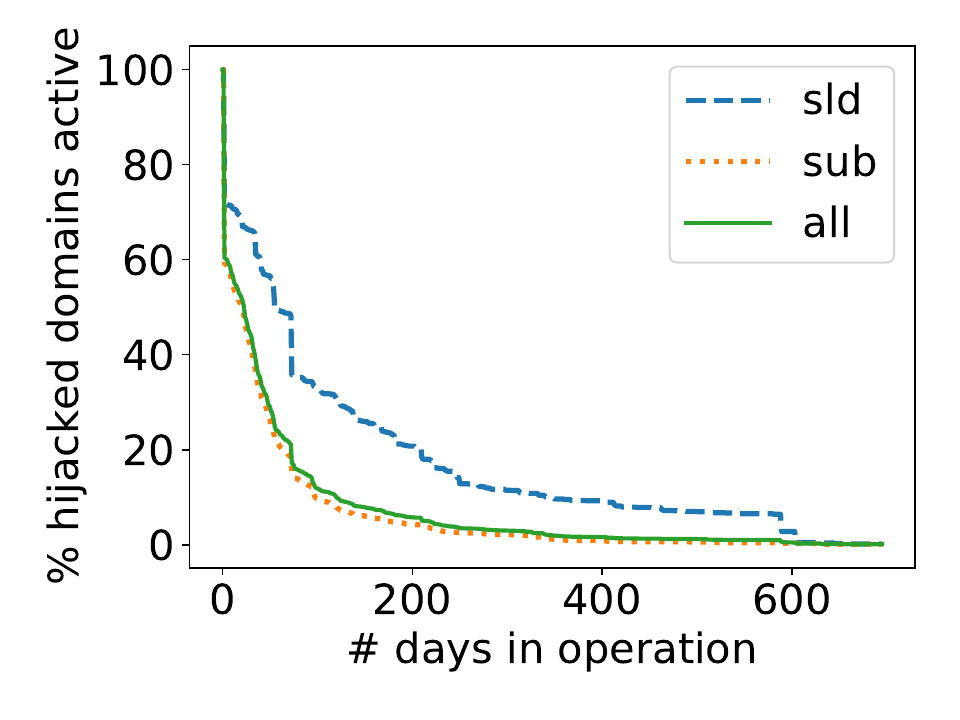}
    \vspace{-25pt}
    \caption{\footnotesize Hijack duration in days.}
    \label{fig:hijack-duration}
  \end{minipage}
  \hfill
  \begin{minipage}[t]{0.2412\textwidth}
    \includegraphics[width=\linewidth,trim=0cm 0cm 0cm 1.3cm, clip]{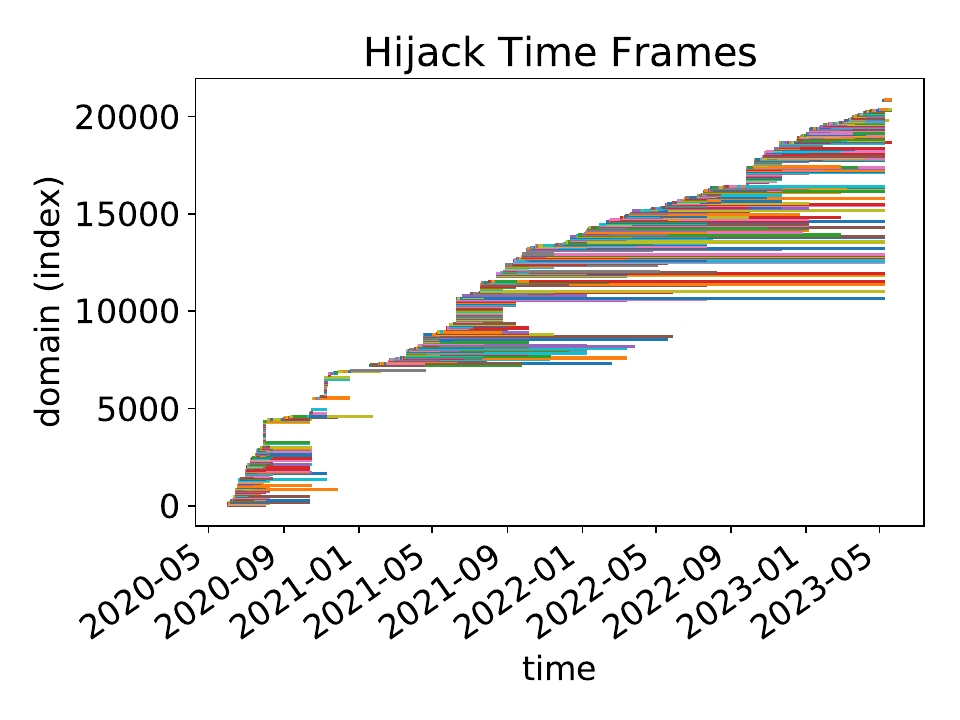}
    \vspace{-25pt}
    \caption{{\footnotesize Hijack time frames.}}
    \label{fig:hijack-timeframes}
  \end{minipage}
  \vspace{-10pt}
\end{figure}

We calculate the approximate lifespan of the abused domains as the difference between the timestamp of the first HTML sample that is recognized as abused and the timestamp of the DNS record that is eventually created by the domain owner to correct the dangling vulnerability. Figure \ref{fig:hijack-duration} shows the lifespan distribution of the domains in our dataset. A large number of abused domain names are removed within 15 days. At the same time, more than $\frac{1}{3}$ of the domains last longer than 65 days, some more than a year. This gives the attacker time to monetize content by exploiting the reputation of the abused domains. Figure \ref{fig:hijack-timeframes} illustrates for each domain the time frame that it was hijacked as a horizontal line from start to end date. The domains are sorted by start date. We see an initial period of hijacks in 2020, followed by a period of relative inactivity in early 2021, and finally a ramping up of activity throughout late 2021, 2022 and 2023. The number of concurrently hijacked domains continuously increases in this period, indicating a growing problem.

\section{Characterization of Abuse}\label{sc:abuse}
In this section we first explain how the attacks with hijacked resources depend on the type of dangling resource that the attacker took over. We then report on different types of attacks we identified that have been launched from the hijacked domains in our dataset. These include SEO, malware distribution, cookie theft and fraudulent certificates.

\subsection{Abuse Depends on Hijacked Resource}
The capabilities of the adversary are dictated by the type of cloud resource used in the hijack (Table \ref{tab:rsc-capabilities}).

\begin{figure}[ht!]
\vspace{-5pt}
    \centering
    \includegraphics[width=1.0\columnwidth]{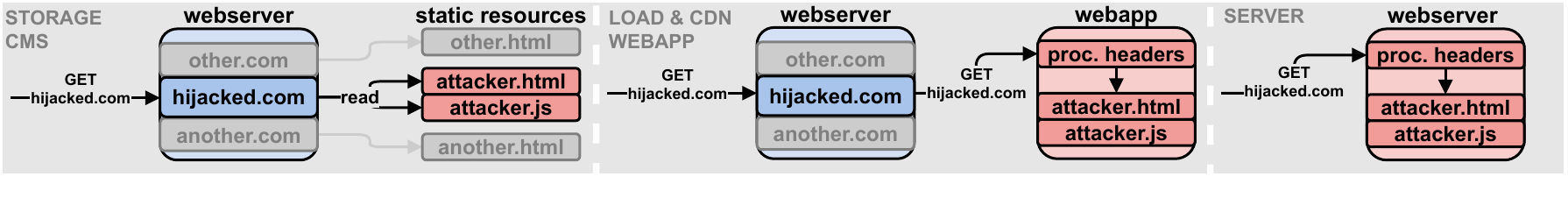}
    \vspace{-20pt}
    \caption{Attacker capabilities based on cloud resource - red indicates attacker-controlled resources, blue indicates hosting-provided resources.}
    \vspace{-5pt}
    \label{fig:hijack-capabilities}
\end{figure}

With a dedicated server resource the attacker can host a full webserver (shown in Figure \ref{fig:hijack-capabilities} on the right). With other resources the hosting provider may host multiple domains on a single logical webserver through virtual hosting and route to the appropriate resource by domain name. Storage resources such as AWS S3 or Content Management Systems (CMS) such as Pantheon allow control only of static content, which is read and returned by the provider's webserver (Figure \ref{fig:hijack-capabilities}, left), whereas with traffic management and web applications the provider's server forwards requests to a specified endpoint, where the requests can be processed in full (Figure \ref{fig:hijack-capabilities}, center). 

In the context of cookie stealing, control of the webserver (Figure \ref{fig:hijack-capabilities}, center \& right) affords the attacker both access to cookies in \texttt{headers} and those available via \texttt{javascript}, thus allowing access to all cookies, whereas control of just the content (Figure \ref{fig:hijack-capabilities}, left) only enables access to cookies accessible via \texttt{javascript}\footnote{In the case of CMS, the use of JS may require the installation of additional plugins. However, since the attacker controls the resource, this is straightforward.}, in other words, cookies without the HttpOnly flag. The former also affords the \texttt{https} capability, required to access cookies with the Secure flag enabled, whereas the latter scenario does not necessarily afford this capability for the hijacked domain. \cite{squarcina2021can} characterized the limitations that configurations on the victim webserver impose on attacks, e.g., the impact of HttpOnly and Secure flags in cookies or that bypassing CSP requires only \texttt{file} and \texttt{html} capabilities, while abuse of CORS, postMessage and domain relaxation also requires the \texttt{javascript} capability. In our model we show that all of these are possible from static hosting resources. However, depending on the configuration of the target, \texttt{https} may be necessary, which generally requires full webserver access to configure a certificate\footnote{Hosting providers may offer a dashboard option to configure a certificate, but this is not the default.}. In addition to the configuration of related domains (shown by \cite{squarcina2021can}), the type of attacks that can be launched with a hijacked domain are also a function of the cloud resource that the attacker controls.

\begin{table}[t!]
\centering
\resizebox{\columnwidth}{!}{
\begin{tabular}{c|c|c|c}
\textbf{Resource}     & \textbf{Function}                      & \textbf{Access} & \textbf{Capabilities}\\ \hline
AWS S3                & Storage                                & \multirow{2}{*}{Static Content} & \multirow{2}{*}{\shortstack[c]{file, content,\\html, javascript$^2$}} \\ \cline{1-2}
Pantheon Site         & CMS                                    &                                 \\ \hline
Netlify               & \multirow{4}{*}{Web App}               & \multirow{8}{*}{Full Webserver} & \multirow{8}{*}{\shortstack[c]{file, content,\\html, javascript,\\headers, https}} \\ \cline{1-1}
Heroku                &                                        &                                 & \\ \cline{1-1}
AWS Elastic Beanstalk &                                        &                                 & \\ \cline{1-1}
Azure Web Application &                                        &                                 & \\ \cline{1-2}
Azure CDN             & \multirow{3}{*}{\shortstack[c]{CDN \&\\Load Balancing}} &                                 & \\ \cline{1-1}
Azure Load Balancer   &                                        &                                 & \\ \cline{1-1}
Cloudflare            &                                        &                                 & \\ \cline{1-2}
Azure VM              & Server                                 &                                 & 
\end{tabular}

}
      \vspace{-5pt}
\caption{Attacker capabilities based on cloud resource}
      \vspace{-15pt}
\label{tab:rsc-capabilities}
\end{table}

\subsection{Generating Traffic}

We find that the main abuse of hijacked, dangling resources is to generate traffic to adversarial services. The attackers exploit the reputation of the hijacked domains to generate page impressions to the content they control to earn money. Once they control the content, sources of income are either advertisements displayed directly on the websites hosted on the hijacked domains or referral (click-through) to another site, where they earn a small amount for each page impression, a higher amount for account registration (Figure \ref{trafficreferral} in Appendix) and even more for money spent.

Attackers use different techniques to generate traffic and increase the click-through rate to the target site that pays for the traffic.
Next we describe the two techniques (SEO and clickjacking) for which we find evidence in our dataset.

\subsubsection{Blackhat Search Engine Optimization (SEO)}

Search Engine Optimization (SEO) is the process of improving a website's visibility in search engine results. Blackhat SEO or spamdexing\footnotemark{}\footnotetext{\url{https://en.wikipedia.org/wiki/Spamdexing}} involves ethically questionable techniques or violates search engine guidelines. We found that 75\% of HTML samples we collected contain some form of (blackhat) SEO. \editB{We determine this by manually examining a sample of 100 HTMLs in a sandboxed environment and then checking which of the other sites contain similar content based on keyword features. Specifically, we found the following techniques:}

\textbf{Cloaking.} The \textit{Japanese Keyword hack}\footnotemark{}\footnotetext{\url{https://web.dev/fixing-the-japanese-keyword-hack/}} is one example of cloaking, a technique where content presented to search engine spiders is different from what is presented to the user. About 1\% of the sites (Figure \ref{fig:content-classification}) featured a large number of randomly generated Japanese pages, which we categorize as the \textit{Japanese Keyword Hack}. If hackers have the ability to add content to a site, they upload a large number of randomly named HTML pages (see Figure \ref{fig:sitemap-file-counts}) with auto-generated Japanese content. These \textit{cloaked} pages are served in parallel to the original site content, but shown only to crawlers, not regular users. Search engines then associate the site and its reputation with the parasitic content. Additional modifications of the \texttt{.htaccess} and \texttt{robots.txt} files in the website's root directory point the crawlers to the generated spam pages and away from the legitimate content. We find that 7.17\% of SEO creates private link networks and utilizes the Japanese Keyword Hack.

\textbf{Private link networks.} Some websites host a large number of files with the sole purpose of linking to other pages and domains , without any valuable content of their own. The reputation of incoming links contributes to the reputation of the target, so hackers create 2-way link networks across pages on hijacked subdomains, exploiting their reputation; we explain this in Section \ref{section-domain-reputation}. In our dataset the abusers uploaded a total of about 500M files with an estimated size of 25.8TB, with an average of 31,810 HTML files per site; statistics are visualized in Figure \ref{fig:sitemap-file-counts}.

\textbf{Doorway pages.} These are low-quality web pages created to rank highly in search results, but link or redirect visitors to a target page that enables monetization. Most of these doorway pages we observed were gambling-related and featured gambling content from Wikipedia. We find that 62.13\% of the SEO uses doorway pages.

\textbf{Keyword stuffing.} This technique optimizes keywords, placing them in the content and keyword meta tag: 41\% of 58353 HTML pages we analyzed contain the keyword meta tag to help make them more discoverable. Table \ref{tab:keyword-counts} shows the top 12 keywords used.

\begin{table}[h!]
\vspace{-3pt}
\renewcommand{\arraystretch}{0.7}
\centering
\scriptsize
  \resizebox{\columnwidth}{!}{%
        \begin{tabular}{c|l|r||c|l|r}
        \textbf{\#} & \textbf{Keyword} & \textbf{Count} & \textbf{\#} & \textbf{Keyword} & \textbf{Count}  \\
        \hline
        1 & slot & 144,108 & 2 & online & 77,669 \\ 
        \hline       
        3 & judi (gambling) & 60,521 & 4 & situs (website) & 35,265 \\
        \hline
        5 & joker123 & 23,630 & 6 & terpercaya (trusted) & 19,407 \\
        \hline        
        7 & gacor (hot streak) & 18,006 & 8 & agen (agent) & 16,939 \\
        \hline
        9 & daftar (register) & 12,881 & 10 & game & 12,113 \\
        \hline
        11 & bola (football) & 11,688 & 12 & pulsa (credit) & 10,467 
        \end{tabular}
    }
    \vspace{-5pt}
  \caption{Top 12 meta tag keywords on content hosted on hijacked domains.}
  \vspace{-5pt}
  \label{tab:keyword-counts}
\end{table}

\subsubsection{Click-Jacking} With this technique an \texttt{onClick} event is inserted early in the event bubbling pipeline to intercept a user's mouse click on a legitimate looking hyperlink and redirect it to a malicious JavaScript function. Through our manual inspection we found this method was used on adult-related pages. Instead of navigating to the indicated page, the user is redirected to another server where ads are served.

\subsubsection{Reputation of Abused Domains}\label{section-domain-reputation} We find that hijacked subdomains are valuable for hackers due to their reputation (or that of the parent domain), which often takes years to establish. Injecting fraudulent content on hijacked subdomains exploits the historic reputation of the parent domain, ranking high in search engine results. Google is the primary source of traffic to websites and accounts for 85-92\% of all search engine traffic for the past 8 years \footnotemark{}\footnotetext{\url {https://www.statista.com/statistics/216573/worldwide-market-share-of-search-engines/}}. As such both legitimate and fraudulent content is optimized to rank as high as possible in search results. Details of Google's ranking algorithm and weights of ranking signals are proprietary and change over time. Nevertheless, domain age, as well as other parameters discussed below, play a role in the ranking. 

\begin{figure}[htbp]
    \centering
    \vspace{-5pt}
    \includegraphics[width=\columnwidth,trim=0cm 0cm 0cm 0.7cm, clip]{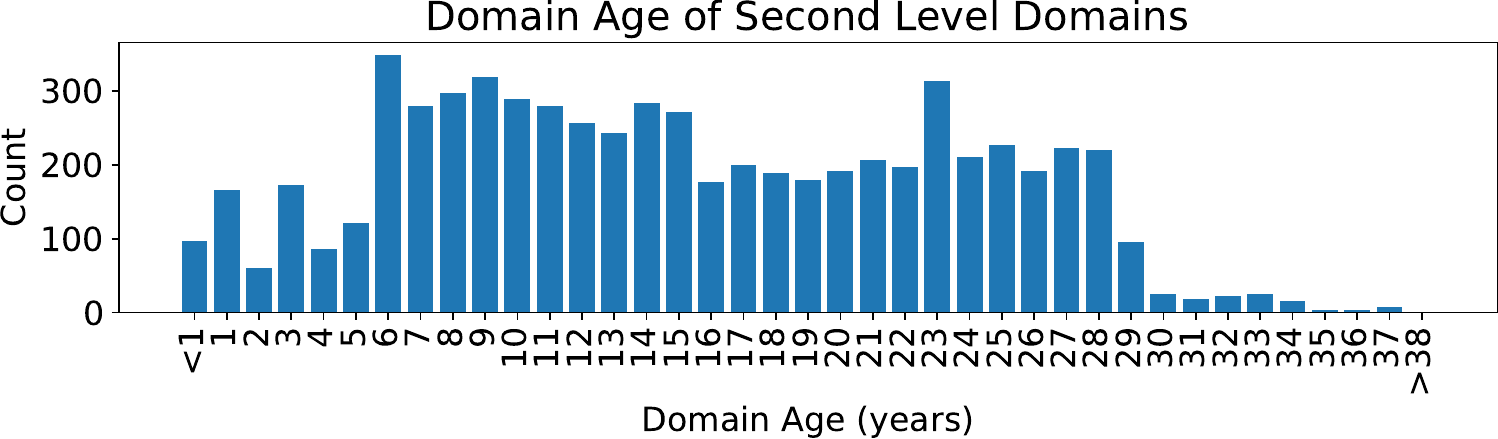}
          \vspace{-20pt}
    \caption{\edit{Domain age based on WHOIS creation date of SLDs.}}
          \vspace{-10pt}
    \label{fig:domain-ages}
\end{figure}

\textbf{Domain age.} \edit{Nearly all} (98.51\%) of the hijacked second-level domains are older than one year \edit{and the vast majority is older than a decade. This is clearly seen in} Figure \ref{fig:domain-ages}\edit{, which} shows the distribution of creation dates \edit{obtained through WHOIS for second-level domains in our dataset. Since subdomains inherit their reputation from the parent, we look at the second-level domain for each subdomain}.

\textbf{Secure transport.} Google also gives preference to sites that are secure. We find that 18.2\% of the (sub)domains in our dataset had valid certificates.

\textbf{Backlinks.} Backlinks play an important role in domain selection by attackers. When other reputable websites link to a domain, Google sees this as a sign of trust and authority. This is especially true if the linking sites are themselves high-reputation domains.

\subsection{More Page Views Mean More Profit} In summary, scammers leverage the reputation inherited from the parent domain to drive search traffic to fraudulent content on the hijacked subdomain. Since search engines are the primary source of traffic, scammers optimize SEO signals to boost their pages' rankings.

In Appendix, Figure \ref{trafficreferral} we show a screenshot of the traffic accounting infrastructure of a gambling site that receives traffic from the hijacked domains. The resources behind the hijacked domains are used to relay traffic to the sign-up screen in Figure \ref{trafficreferral}. In this infrastructure a referral code is passed from the content hosted on a hijacked domain to the gambling website. The website pays for traffic based on the referral ID attached to incoming requests, thereby paying the hijackers for each page view, each account sign-up, and money spent on the site. The presence of the referral ID also suggests that website owners and domain hijackers are two different entities. In this abuse we witness an ecosystem, involving multiple entities creating revenue from the hijacked resources.

In summary, we find referral links, a large focus on SEO, ads and gambling, hijacking of cloud resources exclusively with user-chosen names requiring particularly low effort to hijack, and a lack of differentiation of abuse content across a wide range of SLDs (see Section \ref{sc:attribution}). We view this to be strong evidence for a financial motive as the driver for these hijacks.

\subsection{Malware Distribution \& Flagged Sites}
We find almost no evidence of malware distribution, which was considered to be one of the main threats of dangling records in previous work \cite{squarcina2021can}. \edit{Since this is the first work to find exploitation of dangling domains, there is no comparable dataset in previous work.} We scanned 58,353 samples of HTML index pages for downloadable executable files for all operating systems, to determine, if hijacked domains are directly used for the distribution of malware or other software and which operating systems are targeted.

{\bf VirusTotal executables analysis.} We retrieved 2628 binaries using cURL. Among these were 181 unique Android apps (.apk) and a single Windows executable (.exe). Only 2 EXEs were labeled as Trojans by VirusTotal\footnote{\url{https://www.virustotal.com}}. The vast majority of APKs were gambling apps corresponding to the gambling sites. Our data suggests that the hijacked domains are \textit{not} predominantly used for large-scale malware distribution.

\begin{figure}[ht!]
    \centering
    \vspace{-5pt}
    \includegraphics[width=0.9\columnwidth]{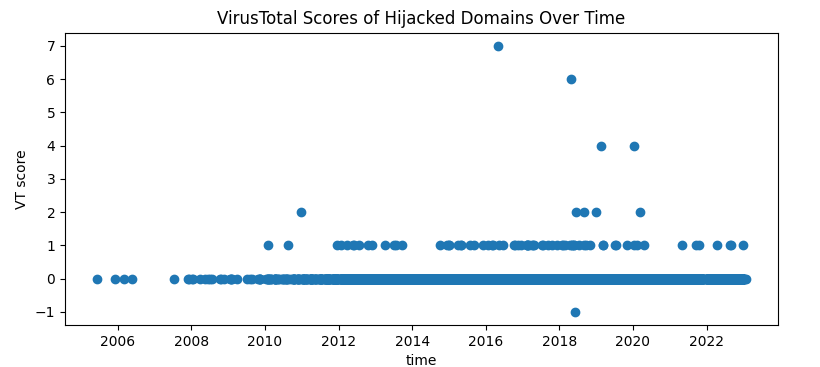}
        \vspace{-5pt}
    \caption{VirusTotal blacklist counts for hijacked domains by date of first certificate issuance.}
    \label{fig:vt-scores}
    \vspace{-15pt}
\end{figure}

{\bf VirusTotal domains analysis.} VirusTotal also allows checking if domains are flagged by various antivirus vendors. We queried VirusTotal for our dataset of hijacked domains: only 135 were flagged by at least one vendor, only 18 by two or more. Figure \ref{fig:vt-scores} plots these values over time (based on the earliest certificate issuance found for that domain)\edit{, suggesting that widespread blacklisting takes at least 2 years}. Whether due to lack of traffic to these domains, the slow inclusion of domains in blacklists or a general disinterest of AV vendors to include hijacked domains, the small percentage of blocked domains suggests that blacklisting does not effectively protect clients from being served content from hijacked domains.

\subsection{Stolen (Authentication) Cookies}\label{sec:cookie-stealing}

As pointed out by \cite{liu2016all,squarcina2021can}, one of the threats of gaining control of a subdomain's content (e.g., access to a CMS) and/or an entire webserver, is access to cookies.
Cookies are used for tracking users across pages via a session identifier, as well as storing authentication tokens which tell the server that a user has previously logged in. 
As a consequence, stealing authentication cookies enables full access to the website as a logged-in user. This sensitive data can be stolen as they visit the legitimate-looking, but adversary-controlled content or webserver on a hijacked subdomain.\\
\indent Cookie policies implemented by browsers are such that a cookie is only sent back to the domain that created it, or a subdomain thereof.
Cookie access depends on the degree of control an adversary has of a hijacked domain. With full server access, they can read all cookies (including authentication), but if they are only able to inject content onto a running webserver (of a CMS), they are only able to read a subset of browser-accessible cookies: cookies with the \texttt{HttpOnly} attribute set to ``false" or not set at all (the default).\\

\indent {\bf Stolen cookies of abused domains.} Since there is no way to tell if cookies are being actively exfiltrated from HTTP headers on the server, we look for stolen authentication cookies for sale. In a collaboration with a threat intelligence organization we identify 83 unique authentication cookies detected in darknet leaks in the timeframe in which the corresponding dangling domains were detected by us as hijacked. These cookies are linked to 3 different hijacked subdomains originating from 53 unique IP addresses. The low number of leaked authentication cookies we found is not surprising, due to their relatively short expiration time.

\subsection{Fraudulent Certificates}\label{sc:fraudulent:cert}

In this section we describe our analysis of the fraudulent certificates we found on hijacked domains.
Due to availability of HTTP-based Domain Validation, hijackers are in a position to obtain a valid certificate, simply by having access to the webserver root.
In fact, many hosting providers, such as Azure, integrate functionality within their dashboards to issue certificates for custom domains that point to the hosting resource. Once a hijacked subdomain is taken over, obtaining a valid certificate is trivial. We can therefore expect hijackers to use certificates, as it allows the use of HTTPS, increasing the efficacy of various attacks. One of the more critical possible abuses for domain hijacks is access to cookies. As detailed in Section \ref{sec:cookie-stealing}, subdomains will typically receive not just cookies set for that subdomain, but those for the parent domains as well. However, if a cookie has the "Secure" flag set, it will only be sent via HTTPS. So attackers looking to steal, e.g., secure authentication cookies would need to setup HTTPS with a valid certificate for the hijacked subdomain. An interested reader is referred to Section \ref{sc:certificates:motivation} in Appendix for explanation on different motivations for obtaining a valid but fraudulent certificate for abused resources.

\subsubsection{Analysis of Certificates on Abused Domains}

Certificate Transparency (CT) logs publicly list all certificates issued by CAs. We use CT to analyze the entire timeline of certificates for all domains in our dataset, looking for anomalies. Across CT history for our dataset we find 24239 single-SAN (single-Subject Alternative Name) certificates and 41877 multi-SAN/wildcard certificates. Since during domain validation hijackers can typically only successfully prove control of a single subdomain that they control, we search the CT history of hijacked subdomains for certificates that contain only a single, non-wildcard subdomain name.

\begin{figure}[t!]
    \centering
    \includegraphics[width=0.9\linewidth]{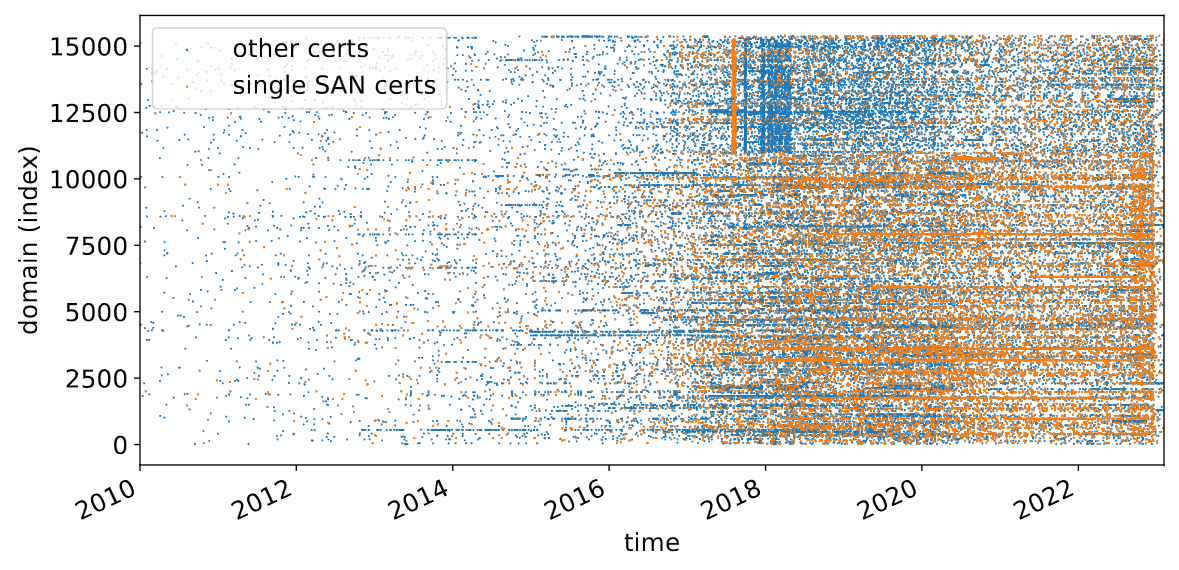}
    \vspace{-10pt}
    \caption{Multi-SAN vs. single-SAN certs issued for hijacked subdomains.}
    \vspace{-15pt}
    \label{fig:single-san}
\end{figure}

We contrast this set with the certificates issued for these subdomains (Figure \ref{fig:single-san}). We see two distinct time frames (2017-07-31 to 2017-08-14 and 2022-09-09 to 2022-12-16) where a significant number of certificates were individually issued across our set of hijacked subdomains. Particularly the former shows a clear anomaly, covering a significant number of "www." subdomains. Taking a closer look at these time frames, we find that the majority of the single-subdomain certificates (95\% and 53\%, respectively, for each time frame) were issued by Let'sEncrypt, suggesting that these were intentionally issued, as opposed to automatically by a hosting provider. A correlation with our dataset of the abused domains shows that the time frames correspond to campaigns to gather vulnerable subdomains. Based on the dates, we trace this kind of activity back to mid 2017, about a year after the first paper on dangling DNS records was published \cite{liu2016all}.

\subsubsection{CAA Records are not Effective}

Hijackers have many options for obtaining a certificate. Cloud providers, such as Azure and AWS, often run their own CAs and provide certificate issuance built into their hosting products, free of charge, which can be leveraged by anyone in control of the hosted resource. Let's Encrypt and ZeroSSL are two additional CAs that provide certificates via domain validation and at no cost.

It has been suggested that domain owners could configure CAA records to allow certificate issuance only by an authorized set of CAs, in order to prevent unauthorized issuance. This is unlikely to be effective, however, since an attacker can simply register an account with one of the authorized CAs and still issue a certificate. This kind of restriction would only be protective if certificate issuance is restricted to a specific \textit{account} with a specific CA ("domain locking"\footnote{https://docs.digicert.com/en/certcentral/manage-certificates/organization-and-domain-management/domain-locking.html\#locking-a-domain}).

In can be argued that, in the case of cyber-criminals gathering large numbers of domains to use for SEO and traffic generation, authorizing only CAs who charge for certificates (i.e., unauthorizing all CAs with free certificates) might disincentivize attackers from issuing certificates at scale, due to an increase in cost. This would, however, necessitate the majority of domain owners switching to a paid CA and setting the appropriate CAA records, in order to function as a deterrent.

This is unlikely to happen due to cost to the legitimate owners. Hence, we find that only 2\% of parent domains (and only 0.2\% of subdomains) have a CAA record set and only 0.4\% (0.01\%) specify a CA without free certificate issuance. We also find that half of these domains still had hijacked subdomains with valid certificates. This suggests that CAA records are not a suitable countermeasure for such attacks.

\subsubsection{Certificate Transparency as Countermeasure}

CT can be leveraged as a much more effective countermeasure than CAA records. Though reactive, rather than preventative, CT monitoring of one's domain via third-party services is a low-to-zero cost, set-and-forget measure to ensure one is notified whenever a certificate is issued for the domain or one of its subdomains. Should an attacker take over even a long-forgotten subdomain and issue a certificate, an alert is triggered and the domain owner is made aware of the hijack, typically within a few hours. However, the effectiveness of detection rests on the attacker's choice to obtain a certificate.

\section{Characterization of Attacker Infrastructure}\label{sc:attribution}
In this section we characterize the infrastructure used in the hijacks in our abuse dataset. We look for indicators from the infrastructure and the user-facing content side. Due to the large number of hijacks we combine manual analysis of a subset of examples with automated keyword-based approaches.

{\bf Hijacked webserver software.} Using the Generator Meta tag, we identified that about 22\% of the 54,325 collected HTML samples were homepages of WordPress blogs. These are custom WordPress installations on cloud.

\begin{figure}
      \centering
   \includegraphics[width=\columnwidth]{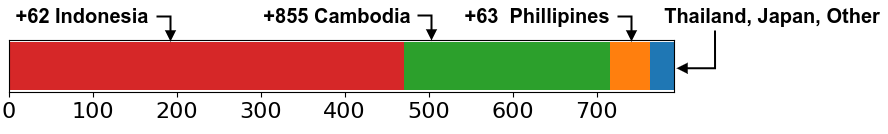}
     \vspace{-20pt}
 \caption{Geo-distribution of phone numbers based on country code.}
 \vspace{-15pt}
    \label{fig:phones_geo}
\end{figure}

{\bf Static identifiers.} Next we considered other backend links present in the HTML samples. We analyzed the \texttt{href} attributes of \texttt{<a>} and \texttt{<link>} tags contained in the HTML documents served by the hijacked domains. We discovered 792 unique phone numbers through WhatsApp links. Based on the country code, we can see that all of them are based in Asia, primarily Indonesia and Cambodia; see Figure \ref{fig:phones_geo}. We also discovered 1,884 unique contacts in the form of Telegram, Twitter, Instagram and Facebook accounts/channels/groups and direct chat IDs, as well as 2,671 unique forwarding links provided by URL shortening services.

 We also discovered 3,553 unique references to IP addresses. Based on WHOIS data, we find that the majority of these IPs belong to various hosting providers (Figure \ref{fig:ip_whois} (a) \edit{in Appendix}). This also matches the concentration of IPs in the US, France and Singapore, based on GeoIP information (Figure \ref{fig:ip_whois} (b) \edit{in Appendix}). The use of cloud hosting does not allow identification of the countries where attackers are operating.

\indent {\bf Clustering attacking infrastructure.} All of the above data points serve as identifiers. If two identifiers appear on the same HTML page, it is reasonable to assume that they are associated with the same operation. Thus, we cluster these data points based on how many HTML pages of hijacked domains each pair appears on. The resulting relationships between identifiers are visualized as a network graph in Figure \ref{fig:group_clustering} in the Appendix. The network graph provides a broad overview. To delineate concrete groupings we find all nodes that are connected through some path in the network graph. This is accomplished by successive hierarchical clustering, plotted in Figure \ref{fig:group_dendrogram} in Appendix.

\indent Each tick on the x-axis represents an identifier, while the y-axis indicates the distance (on a scale of 0 to 1) at which identifiers are grouped. The minimum distance of 0 indicates that a pair of identifiers are associated with an identical set of hijacked domains and the maximum distance of 1 indicates that two identifiers share no domains at all. The dendrogram displays all groupings up to a cutoff point at a distance of 0.95 (since at the end of the hierarchical clustering process all nodes are merged into one). The cutoff is chosen to achieve the maximal degree of grouping. This is warranted, since the probability of an identifier appearing on two domains by coincidence is very small. Groupings are delineated by color, so all adjacent vertical lines of the same color point to identifiers that have been grouped; the largest grouping is displayed in gray on the right side of the dendrogram. Long, single-color vertical lines indicate identifiers that do not share any domains with other identifiers and thus cannot be grouped.

\begin{figure}[t]
\centering
    \includegraphics[width=0.9\columnwidth]{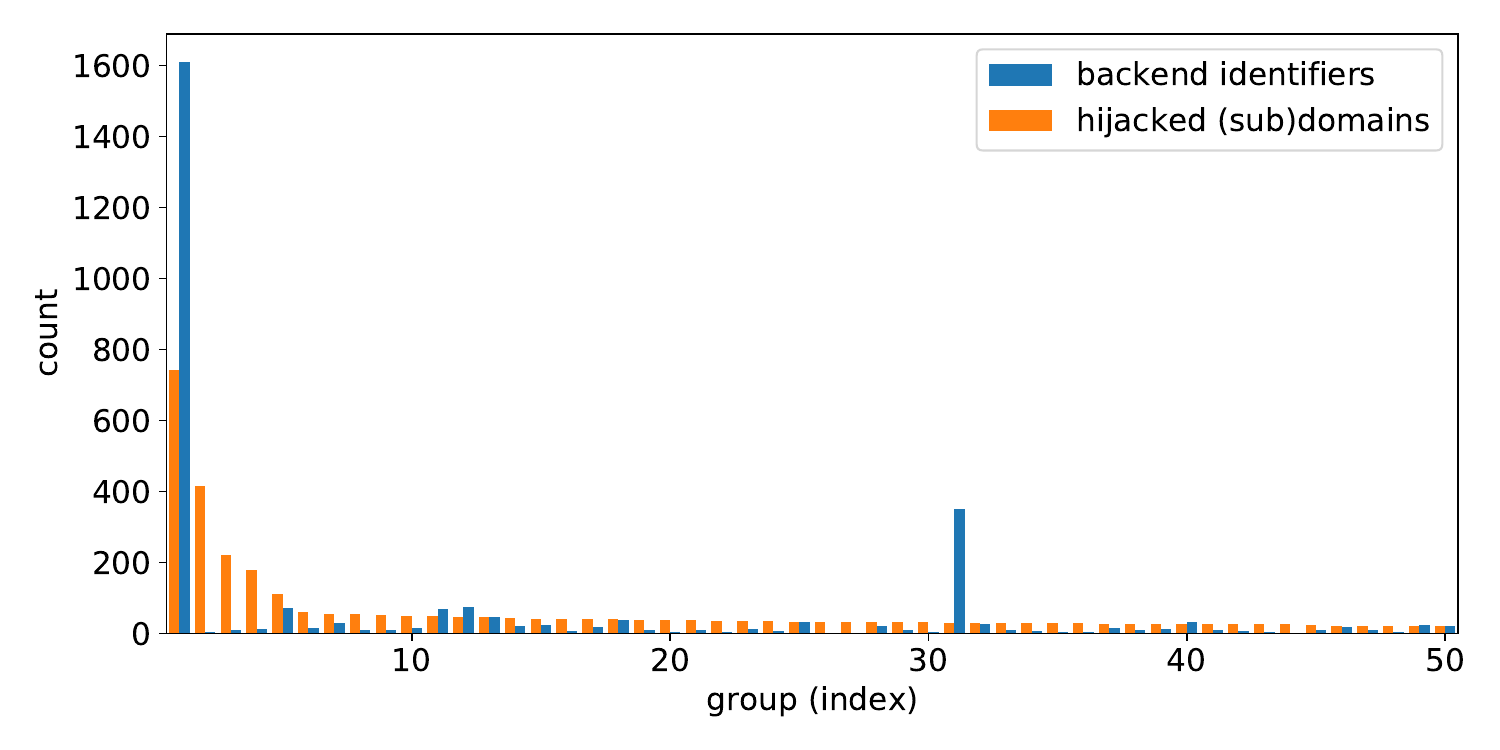}
    \vspace{-10pt}
    \caption{Top 50 clusters sorted by number of hijacked domains.}
    \vspace{-15pt}
    \label{fig:cluster_groups_sorted}
\end{figure}

\indent The hierarchical clustering results in 1,798 clusters, with the vast majority consisting of 1 or 2 identifiers which could not be linked to any others. For the largest grouping, however, it was possible to tie together 1,609 identifiers, associated with 743 hijacked domains. The four next-largest clusters contain 414, 222, 179 and 112 domains, respectively. Figure \ref{fig:cluster_groups_sorted} shows this long-tailed clustering result for the top 50 clusters, sorted by the number of hijacked domains in each cluster. The identifiers cover 8,489 (\textasciitilde$\frac{1}{3}$) of the hijacked domains and all are associated with Indonesian gambling.

\indent While the clustering still leaves the possibility for a large number of actors involved, it ties large sets of domains together, showing that at least some actors collect a wide range of diverse domains in a coordinated effort. These are then homogeneously used for the same purpose of referring traffic or manipulating search rankings. The lack of differentiation in how the hijacked domains are abused suggests an attempt at maximizing profit by maximizing the number of domains recruited for a campaign, as opposed to specific targeting of individual domains, e.g., for political reasons.

\indent \edit{Our analyses do not yet allow conclusions as to the noticeable bias of observed page content, linked infrastructure, discovered APKs and phone numbers toward Indonesia. However, because our discovery of hijacks begins with a set of domains based on global organizations (e.g., Alexa 1M, Fortune 1000, etc.), there is little reason to assume a biased dataset. Instead, we see a possible explanation in the population size (4th largest in the world) and strict illegality of gambling in Indonesia, leading to a prevalence of online gambling and a need to advertise it through illicit means.}

\section{Conclusions}\label{sc:conc}
Although the threat of take-over of dangling resources in the cloud was explored, there was no evidence in research of real-life abuse. We explore this question empirically with a longitudinal three-year analysis of cloud-hosted resources. Our two key contributions are a methodology for detecting abuse and a longitudinal dataset of abused resources at scale. Our research shows that the abuse of dangling resources on cloud platforms is a real problem that affects a large number of victims in popular and established organizations across different sectors. Our methodology and the findings provide a feasibility proof for identifying abused resources without assuming control over the cloud platform.

Based on our analysis we derive lessons for improving visibility of abuse and developing countermeasures. The hijacks we found show that the attackers target released resources that (1) are cheap and (2) can be directly determined by entering freetext, while avoiding resources that are expensive and require effort to obtain, such as the lottery-based IP assignment from a pool of IP addresses.
Therefore, as an easy-to-deploy mitigation we recommend that cloud platforms either do not allow user-created resource names to be publicly visible (e.g., through DNS records) and/or disallow the re-registration of recently released resource names. We also recommend, similarly to previous work, to purge stale DNS records. In addition, cloud platforms should keep track of released resources using our methodology and alert owners of registered domains about changes to the content or sitemap. Since we observe that attackers issue certificates for hijacked domains, we recommend that cloud providers also monitor CT logs for unusual patterns across domains hosted on their platforms to help detect potential large-scale abuse campaigns.

Finally, we point out that although our work focuses on resources on cloud platforms, our results can be used to identify abuse in other third-party services. For instance, while Content Management Systems (CMS) like Wordpress are not included in our dataset, we expect a large number of hijacks of \texttt{[freetext].wordpress.com} subdomains, since Wordpress also implements freetext subdomain registration for its blogs.

\section*{Acknowledgements}
This work has been co-funded by the German Federal Ministry of Education and Research and the Hessen State Ministry for Higher Education, Research and Arts within their joint support of the National Research Center for Applied Cybersecurity ATHENE and by the Deutsche Forschungsgemeinschaft (DFG, German Research Foundation) SFB~1119.

\balance
\newpage

\bibliographystyle{plain}
\bibliography{sec,NetSec}

\appendix
\section{Appendix}
\subsection{Cloud Suffixes}\label{sc:cloud:suffixes}
We compiled the list of cloud suffixes using the following sources:\\
{\footnotesize
\path{https://docs.aws.amazon.com/general/latest/gr/rande.html}\\
\path{https://learn.microsoft.com/en-us/azure/security/fundamentals/azure-domains}\\ \path{https://docs.netlify.com/domains-https/custom-domains/}\\
\path{https://docs.pantheon.io/guides/domains/platform-domains}\\
\path{https://devcenter.heroku.com/articles/custom-domains}\\
\path{https://infogalactic.com/info/List_of_Google_domains}.}\\

\edit{Cloud IP ranges were obtained from these provider published sources:}\\
\path{https://ip-ranges.amazonaws.com/ip-ranges.json}\\
\path{https://www.microsoft.com/en-us/download/details.aspx?id=56519}\\
\path{https://www.gstatic.com/ipranges/cloud.json}\\

\edit{Based on recent market share data, the AWS, Azure and Google clouds cover 65\% of the hosting market, with the rest split across a long tail of providers. Our cloud identification is therefore not 100\% complete, but covers a large majority share of the cloud market.}

\subsection{Motivations for Obtaining a Certificate for a Hijacked Domain}\label{sc:certificates:motivation}

{\bf Browser Warnings.} Most popular browsers display a warning when attempting to connect to a domain with a self-signed, expired or otherwise invalid certificate. Some end-users might ignore this warning, but many won't. Obtaining and using a valid certificate would remove this barrier and thus increase traffic to the site.

Similarly, browser UI indicators (i.e. green lock icon) increase the users' trust in the legitimacy of the site. It has been shown that phishing sites using HTTPS are more effective \cite{kim2021phishing}. Thus, a certificate for the hijacked site likely serves to increase user interaction.

{\bf SEO.} The primary type of abuse seen on hijacked domains is SEO spam. A key parameter in search engine rankings is the use of HTTPS. Sites that don't use HTTPS are typically ranked lower. Thus, a certificate would help boost the efficacy of SEO spam.

{\bf HSTS.} If a hijacked subdomain (or its parent) has previously added the domain to the HSTS list of visiting clients, these clients will only connect to this domain via HTTPS in the future (until expiration of the HSTS setting). Hijackers wishing to capture this traffic, will need to serve HTTPS connections and will thus require a certificate. We queried 1,323 parent domains from our hijacked dataset and found over 16\% of non-error responses contained an HSTS header.

{\bf Secure Cookies.} One of the more critical possible abuses for domain hijacks is access to cookies. As detailed in Section \ref{sec:cookie-stealing}, subdomains will typically receive not just cookies set for that subdomain, but those for the parent domains as well. However, if a cookie has the "Secure" flag set, it will only be sent via HTTPS. So attackers looking to steal, e.g., secure authentication cookies would need to setup HTTPS with a valid certificate for the hijacked subdomain.

Furthermore, the SameSite flag differentiates HTTPS and HTTP origins. In particular, SameSite=None requires Secure=True, increasing the likelihood that a domain uses secure authentication cookies\footnote{https://developer.mozilla.org/en-US/docs/Web/HTTP/Headers/Set-Cookie/SameSite}.

\begin{figure}[th!]
\centering
\includegraphics[width=0.5\textwidth]{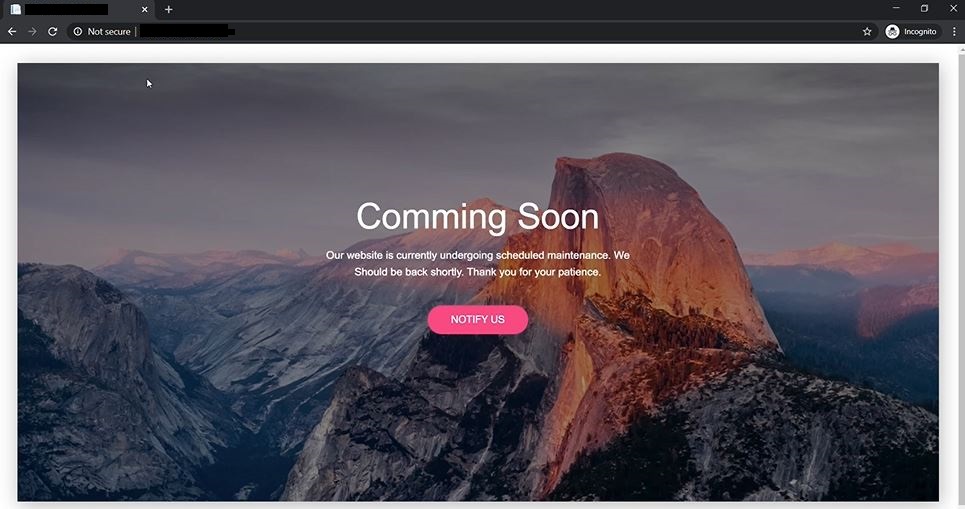}
\caption{Main page for a Fortune 500's abused site with an error message (May 2020).}
\label{abuse-2020-05-censured}
\end{figure}

\begin{figure}[t!]
\centering
\includegraphics[width=\linewidth,trim=0cm 0.5cm 0cm 0cm, clip]{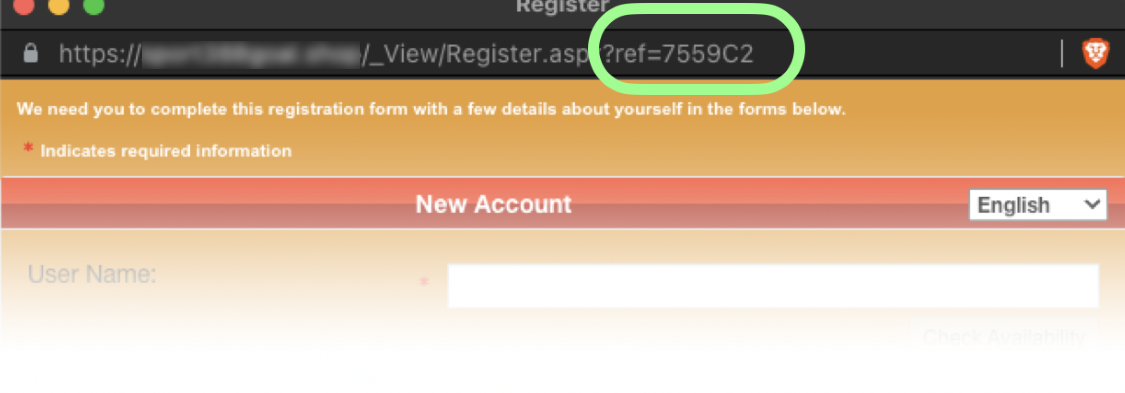}
\caption{Referral code is passed for traffic accounting.}\label{trafficreferral}
\end{figure}

\begin{figure}[t!]
    \centering
    \includegraphics[angle=-90,width=\columnwidth]{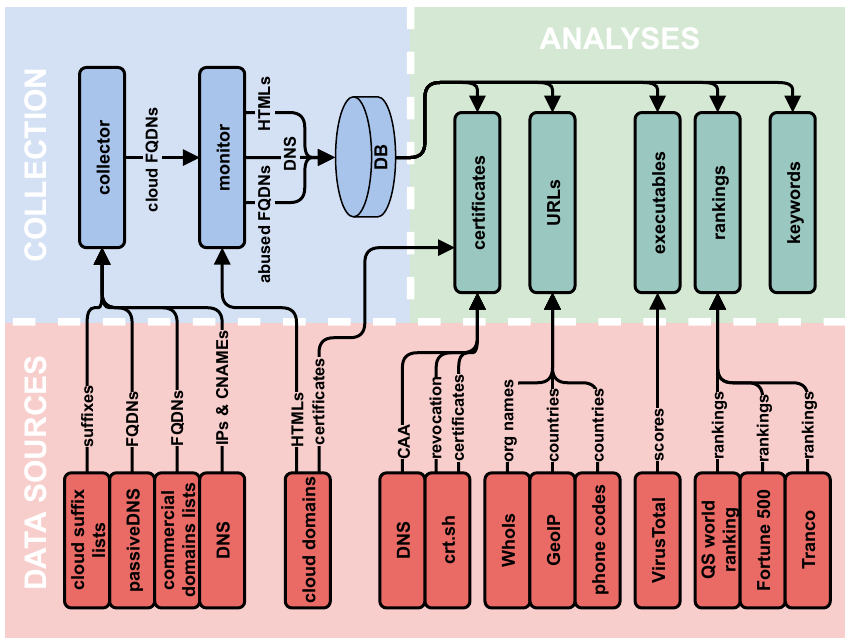}
    \caption{Overview of data collection and analysis process.}
    \label{fig:data-flow}
\end{figure}

\begin{figure}
    \centering
   \includegraphics[width=\columnwidth]{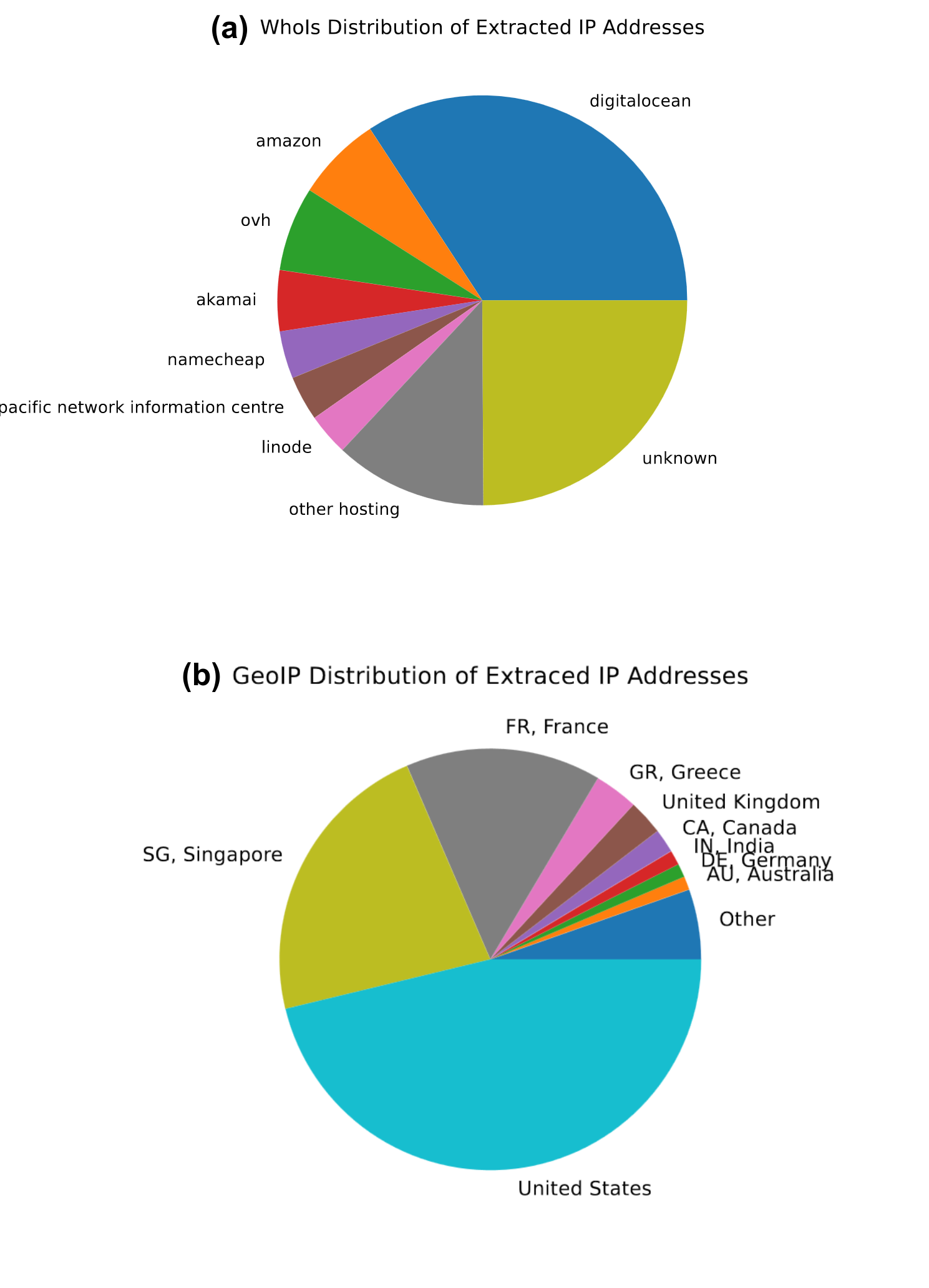}
    \caption{\textbf{(a)} Organizations associated with extracted IPs based on WHOIS data. \textbf{(b)} Geographical distribution of extracted IPs based on GeoIP data.}
    \label{fig:ip_whois}
\end{figure}

\begin{figure*}[t!]
    \centering
    \includegraphics[width=\textwidth]{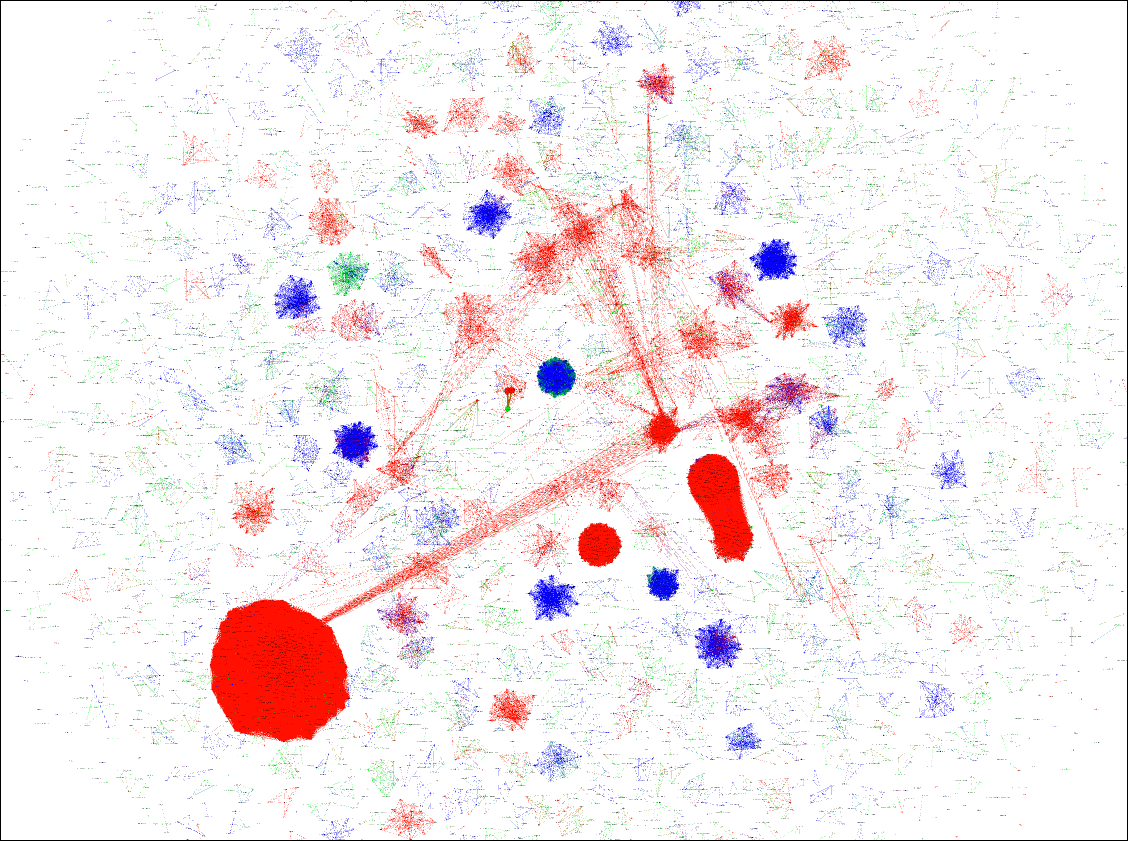}
    \caption{Clustering of extracted identifiers - IPs (red), contact information such as phone numbers, social media channels and chat links (green), and URL shortener links (blue). Node size indicates the number of hijacked domains associated with the identifier, edge thickness indicates the number of shared domains between a pair of identifiers.}
    \label{fig:group_clustering}
\end{figure*}

\begin{table}[t!]
\renewcommand{\arraystretch}{0.7}
\centering
\scriptsize
\begin{tabular}{l|c|l|c}
\hline
\textbf{TLD} & \textbf{Count} & \textbf{TLD} & \textbf{Count} \\
\hline
1. com & 12942 & 7. de & 758 \\
2. org & 1069 & 8. edu & 414 \\
3. net & 996 & 9. ca & 398 \\
4. uk & 758 & 10. nl & 207\\
5. au & 414 &  11. jp & 183 \\
6. br & 398 & 12. co & 156\\
\end{tabular}
\caption{Top 12 Top Level Domains (from a total of 218) and their counts.}
\label{table:tld_counts}
\end{table}

\begin{figure*}[t!]
    \centering
    \includegraphics[width=\textwidth]{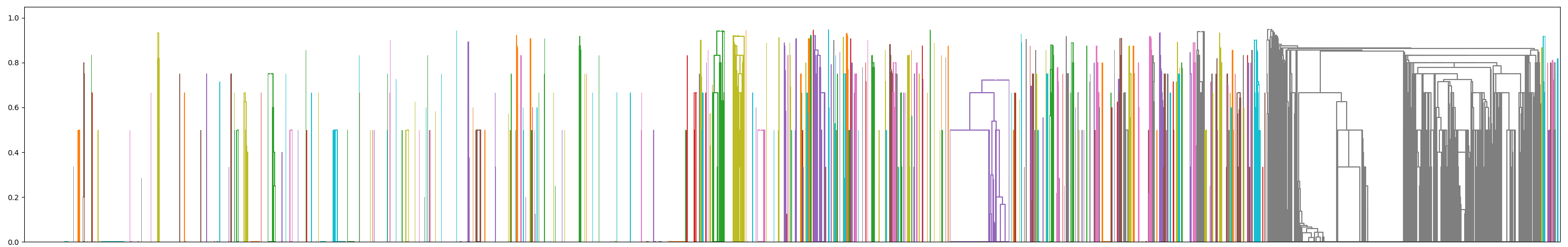}
    \caption{Dendrogram of identifier clustering.}
    \label{fig:group_dendrogram}
\end{figure*}

\begin{figure*}[t!]
    \centering
    \includegraphics[page=1,width=\textwidth]{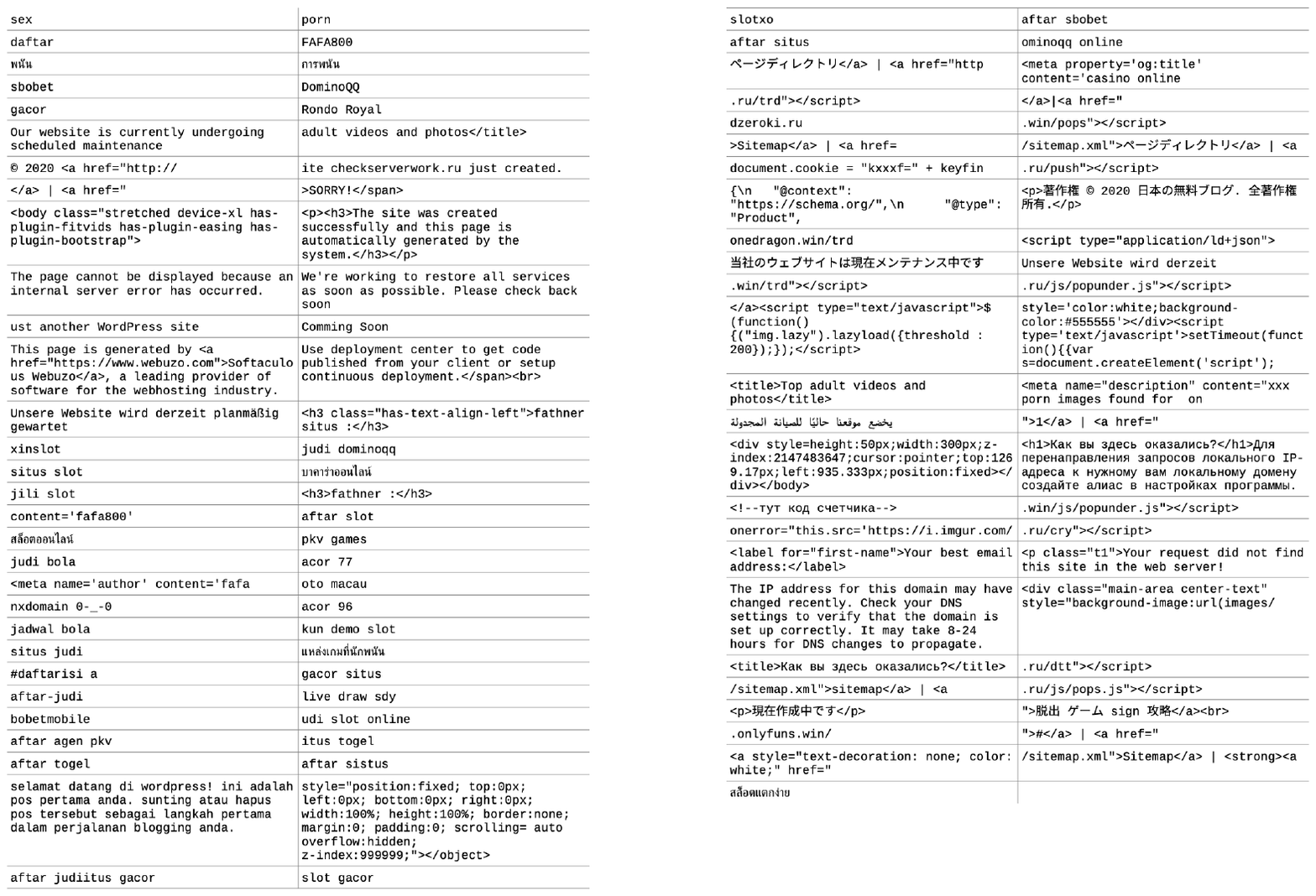}
    \caption{List of extracted keywords and code fragments.}
    \label{fig:keyword-table}
\end{figure*}

\end{document}